\documentclass[twocolumn]{aastex63}

\received{}
\revised{}
\accepted{}
\submitjournal{ApJ}

\shorttitle{Sun-as-a-star Observations of Transiting ARs}
\shortauthors{Toriumi et al.}
\begin{document}

\title{Sun-as-a-star Spectral Irradiance Observations of Transiting Active Regions}

\correspondingauthor{Shin Toriumi}
\email{toriumi.shin@jaxa.jp}

\author[0000-0002-1276-2403]{Shin Toriumi}
\affiliation{Institute of Space and Astronautical Science, Japan Aerospace Exploration Agency, 3-1-1 Yoshinodai, Chuo-ku, Sagamihara, Kanagawa 252-5210, Japan}

\author[0000-0003-4452-0588]{Vladimir S. Airapetian}
\affiliation{Sellers Exoplanetary Environments Collaboration, NASA Goddard Space Flight Center, Greenbelt, USA}
\affiliation{Department of Physics, American University, Washington DC, USA}

\author[0000-0001-5685-1283]{Hugh S. Hudson}
\affiliation{School of Physics and Astronomy, University of Glasgow, Glasgow, UK}
\affiliation{Space Sciences Laboratory, University of California at Berkeley, Berkeley, USA}

\author[0000-0002-6010-8182]{Carolus J. Schrijver}
\affiliation{Lockheed Martin Solar and Astrophysics Laboratory, 3251 Hanover Street, Building/252, Palo Alto, CA 94304, USA}

\author[0000-0003-2110-9753]{Mark C. M. Cheung}
\affiliation{Lockheed Martin Solar and Astrophysics Laboratory, 3251 Hanover Street, Building/252, Palo Alto, CA 94304, USA}

\author[0000-0002-6338-0691]{Marc L. DeRosa}
\affiliation{Lockheed Martin Solar and Astrophysics Laboratory, 3251 Hanover Street, Building/252, Palo Alto, CA 94304, USA}

%% Note that the \and command from previous versions of AASTeX is now
%% depreciated in this version as it is no longer necessary. AASTeX 
%% automatically takes care of all commas and "and"s between authors names.

%% AASTeX 6.3 has the new \collaboration and \nocollaboration commands to
%% provide the collaboration status of a group of authors. These commands 
%% can be used either before or after the list of corresponding authors. The
%% argument for \collaboration is the collaboration identifier. Authors are
%% encouraged to surround collaboration identifiers with ()s. The 
%% \nocollaboration command takes no argument and exists to indicate that
%% the nearby authors are not part of surrounding collaborations.

%% Mark off the abstract in the ``abstract'' environment. 
\begin{abstract}
Major solar flares are prone to occur in active region atmospheres associated with large, complex, dynamically-evolving sunspots. This points to the importance of monitoring the evolution of starspots, not only in visible but also in ultra violet (UV) and X-rays, in understanding the origin and occurrence of stellar flares. To this end, we perform spectral irradiance analysis on different types of transiting solar active regions by using a variety of full-disk synoptic observations. The target events are an isolated sunspot, spotless plage, and emerging flux in prolonged quiet-Sun conditions selected from the past decade. We find that the visible continuum and total solar irradiance become darkened when the spot is at the central meridian, whereas it is bright near the solar limb; UV bands sensitive to the chromosphere correlate well with the variation of total unsigned magnetic flux in the photosphere; amplitudes of EUV and soft X-ray increase with the characteristic temperature, whose light curves are flat-topped due to their sensitivity to the optically thin corona; the transiting spotless plage does not show the darkening in the visible irradiance, while the emerging flux produces an asymmetry in all light curves about the central meridian. The multi-wavelength sun-as-a-star study described here indicates that such time lags between the coronal and photospheric light curves have the potential to probe the extent of coronal magnetic fields above the starspots. In addition, EUV wavelengths that are sensitive to the transition-region temperature sometimes show anti-phased variations, which may be used for diagnosing plasmas around starspots.
\end{abstract}

%% Keywords should appear after the \end{abstract} command. 
%% See the online documentation for the full list of available subject
%% keywords and the rules for their use.
\keywords{Solar spectral irradiance --- Sunspots --- Solar active regions --- Solar flares  --- Solar analogs --- Starspots --- Stellar flares --- Time domain astronomy}

%% From the front matter, we move on to the body of the paper.
%% Sections are demarcated by \section and \subsection, respectively.
%% Observe the use of the LaTeX \label
%% command after the \subsection to give a symbolic KEY to the
%% subsection for cross-referencing in a \ref command.
%% You can use LaTeX's \ref and \label commands to keep track of
%% cross-references to sections, equations, tables, and figures.
%% That way, if you change the order of any elements, LaTeX will
%% automatically renumber them.
%%
%% We recommend that authors also use the natbib \citep
%% and \citet commands to identify citations.  The citations are
%% tied to the reference list via symbolic KEYs. The KEY corresponds
%% to the KEY in the \bibitem in the reference list below. 

\section{Introduction}\label{sec:intro}

The most prominent manifestations of magnetic activity on the Sun are active regions (ARs), which contain sunspots \citep{2003A&ARv..11..153S}, and whose evolution drives solar flares having energies of up to $10^{32}$ erg \citep{2011SSRv..159...19F,2011LRSP....8....6S,2017LRSP...14....2B}. Recent {\it Kepler}, {\it TESS}, and ground-based observations reveal that similar but much more energetic superflares, with bolometric energies of $10^{33}$ to $10^{36}$ erg, can happen on slowly-rotating stars like the Sun \citep{2012Natur.485..478M,2013ApJS..209....5S,2019ApJ...876...58N}. From the consistency of the observed statistical trends between the solar and stellar flares, it is speculated that these two phenomena share common driving mechanism, i.e., the releasing of free magnetic energy in association with magnetic reconnection, probably in the corona above the spots.

From the history of solar flare observation, it is known that major flares tend to emanate from regions associated with large, complex-shaped, rapidly-developing sunspots \citep{2009AdSpR..43..739S,2013ApJ...773..128T,2017ApJ...834...56T,2019LRSP...16....3T}. Therefore, to understand the origin and occurrence of stellar flares, it is important to investigate the size, structure, and temporal evolution of starspots. Although it is still difficult to spatially resolve starspots to examine their morphology and complexity \citep{2005LRSP....2....8B,2009A&ARv..17..251S}, visible photometric observations show that superflare stars harbor starspots or starspot conglomerates of much larger sizes than the sunspots \citep{2013ApJ...771..127N}. The evolution of starspots, in particular their growth and decay rates, has been investigated on collections of solar-type stars \citep{2019ApJ...871..187N,2020ApJ...891..103N}.

In regard to solar ARs, while the visible is sensitive to the photosphere, we can probe the atmospheres of different temperatures in other wavelength bands. For instance, the physical state of the solar corona is diagnosed using extreme ultra violet (EUV) and X-ray observables. In UV of longer wavelengths, we can investigate the chromosphere. Therefore, by conducting multi-wavelength stellar observations, we may probe not only the photosphere but also the corona, where the magnetic field stores and releases the energy through stellar flares. Moreover, ionizing radiation from EUV and X-ray photons may even determine the habitability of exoplanets by controlling the ionization states of exoplanetary atmospheres, which emphasizes the importance of simultaneous observations in these wavelengths \citep{2020IJAsB..19..136A}.

Ever since the total solar irradiance (TSI) monitoring from the space commenced in the late 1970s, it is known that when a sunspot transits the solar disk, the TSI is reduced by an amount that is roughly proportional to the spot area \citep[e.g.,][]{1981Sci...211..700W,1982SoPh...76..211H}. When the spot is near the limb, on contrary, the TSI is enhanced due to the excess emission of faculae because the hot walls of the faculae, which are the vertical magnetic flux tubes in the photosphere, become increasingly visible to the observer \citep{1976SoPh...50..269S,2013ARA&A..51..311S}, as numerically modeled by, e.g., \citet{2004ApJ...607L..59K} and \citet{2004ApJ...610L.137C}. \citet{2004SoPh..222....1Z} analyzed the transit of emerging flux and found an asymmetric light curve, in a search for time-dependent correlation of spots and faculae.

However, up to now, there have been few solar spectral irradiance (i.e., multi-wavelength) studies of transiting sunspots. One such exception is \citet{2004GeoRL..3110802W}, who report that when the huge sunspot group AR 10486 (the maximum spot area being 2610 MSH\footnote{Millionths of the solar hemisphere, which is equivalent to $3\times 10^{6}\ {\rm km}^{2}$.}) crossed the disk, the TSI reduced by 0.34\% ($=3400$ ppm) while the irradiances of \ion{Mg}{2} k (2796~{\AA}) and \ion{H}{1} Ly-$\alpha$ (1216~{\AA}) increased by 20\% and 25\%, respectively.

%Probably there are even fewer multi-wavelength observations of transiting starspots because of the strong UV/EUV extinction by interstellar medium.
On other stars, there are even fewer multi-wavelength observations of transiting starspots. Such studies are hampered because of strong UV/EUV extinction by the interstellar medium and the and the low sensitivity and general lack of long, continuous observations by existing missions. In fact, many of the UV long-term monitoring campaigns have been for active binary systems (RS CVn or BY Dra-type binaries) by {\it IUE}, {\it EUVE}, and {\it FUSE} \citep[e.g.,][]{1983ApJ...267..232S,1992ApJ...391..760S,2006ASPC..348..269R}. In X-rays, rotational modulation has been studied on rapidly-rotating young Suns by {\it XMM-Newton} and {\it Chandra} \citep[][and references therein]{2003A&A...408.1087S,2005ApJS..160..450F,2007LRSP....4....3G}, however investigations into the rotational modulation of slowly-rotating, solar-like stars remain difficult.
Nevertheless, several space UV missions are currently planned, such as {\it ESCAPE} \citep{2019SPIE11118E..08F} and {\it WSO-UV} \citep{2014Ap&SS.354..155S}, and they may provide new and better opportunities to realize the spectral irradiance monitoring of starspots.

In this study, we investigate the sun-as-a-star light curves for different types of transiting ARs at various wavelengths from visible, UV, EUV, to soft X-rays as well as the photospheric total magnetic flux, with the aim of characterizing the evolution of stellar ARs in these wave bands. In particular, we target an isolated sunspot, spotless plage, and emerging flux that appear during prolonged quiet-Sun (QS) conditions where few to almost no competing sunspots or plages are present. From the resulting observables, we discuss ways to exploit stellar light curves for determining the characteristics of stellar ARs.

It should be noted that the applicability of the results to stars whose activity is substantially different from the Sun is not warranted and the comparison is limited to Sun-like stars (slowly-rotating G-type dwarfs) because, for instance, thermal structures of the corona of young, fast-rotating stars are likely to differ from that of the Sun. Nevertheless, through the present study, we may learn how transiting stellar ARs impact the light curves of different wavelengths and how we can obtain a clue to understand the properties of such ARs.

\begin{deluxetable*}{cccccccc}
%\tablenum{1}
\tablecaption{List of Events Analyzed in This Study\label{tab:events}}
\tablewidth{0pt}
\tablehead{
\colhead{Event} & \colhead{Transiting object} & \multicolumn2c{Start} & \multicolumn2c{Central meridian} & \multicolumn2c{End}
}
%\decimalcolnumbers
\startdata
QS & \nodata & 2019-Nov-27 & 00:00 & 2019-Dec-07 & 00:00 & 2019-Dec-17 & 00:00\\
Sunspot & AR 12699 & 2018-Feb-01 & 00:00 & 2018-Feb-11 & 00:00 & 2018-Feb-21 & 00:00\\
Plage & Return of AR 12713 & 2018-Jul-04 & 18:00 & 2018-Jul-14 & 18:00 & 2018-Jul-24 & 18:00\\
Emerging flux & AR 12733 & 2019-Jan-14 & 13:00 & 2019-Jan-24 & 13:00 & 2019-Feb-03 & 13:00\\
\enddata
\tablecomments{All dates and times are in UT.}
\end{deluxetable*}

The rest of the paper is organized as follows. In Section \ref{sec:analysis}, we describe the event selection and the observational data we analyzed. The resulting light curves are presented in Section \ref{sec:results}, whereas the detailed investigations are performed in Section \ref{sec:detailed}. Finally, we summarize and discuss the results in Section \ref{sec:summary}.

\section{Data Analysis}\label{sec:analysis}

In this study, we analyzed solar data for 20-day periods during which QS conditions continued or only a single isolated sunspot, plage or flux emergence region made a transit across the solar disk. We searched for such events by browsing the monthly {\it SDO}/HMI magnetogram animations from May 2010 through April 2020. However, ideal events were found only during the solar minimum between Solar Cycles 24 and 25. The time intervals for each case along with the reference QS period are summarized in Table \ref{tab:events}.

\begin{figure*}
\begin{center}
\includegraphics[width=0.9\textwidth]{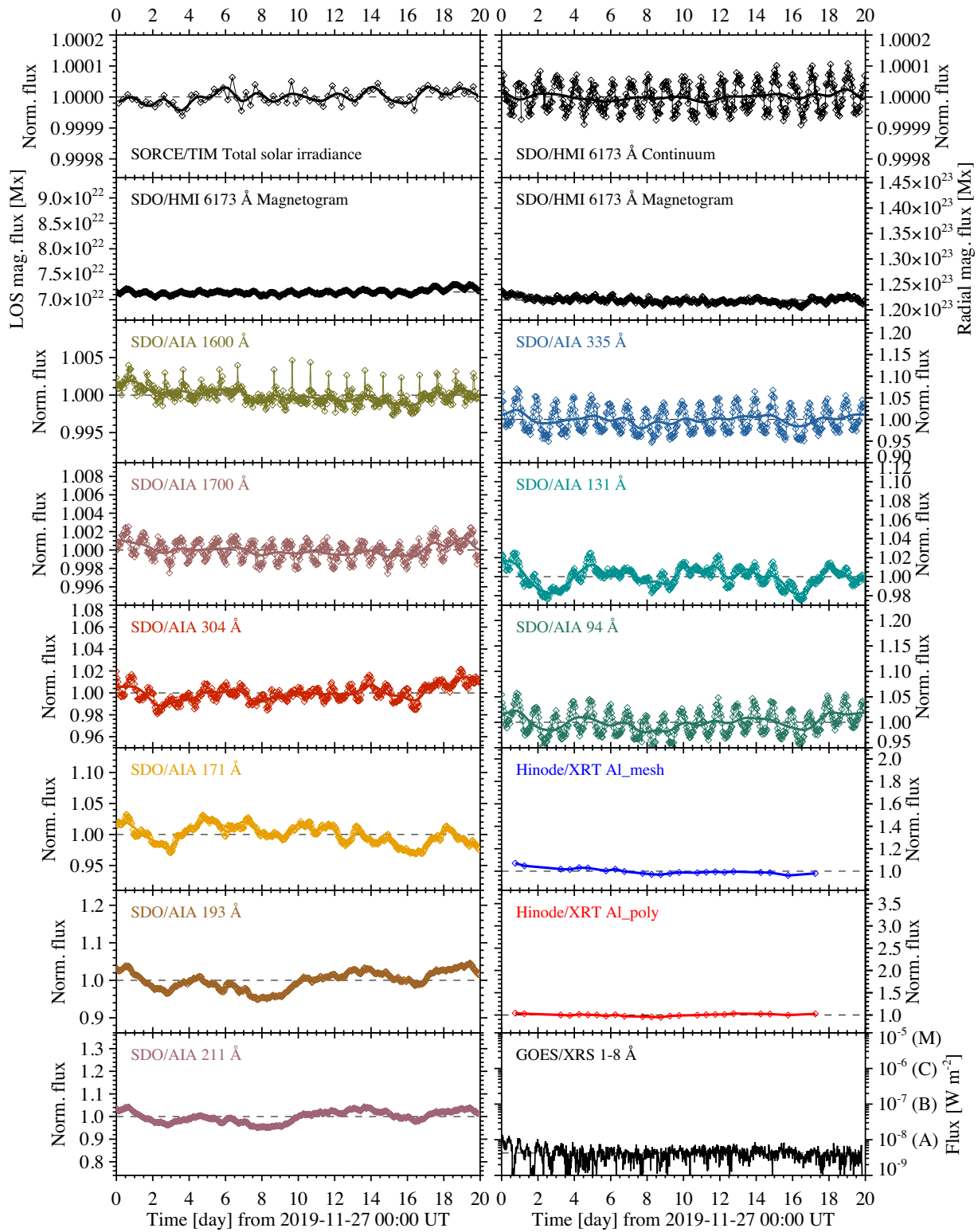}
\end{center}
\caption{Collection of the disk-integrated light curves for the QS over the 20 days starting at 00:00 UT on 2019 November 27. Diamonds represent the original data and, except for the {\it Hinode}/XRT and {\it GOES}/XRS, thick solid lines are the 36-hr smoothed curves (see main text for the details). In most light curves, the fluxes are normalized by the mean values of the smoothed curves. Dashed horizontal lines indicate the level where the normalized flux is unity. Because the variation of {\it GOES}/XRS for transit events is significantly larger than those of the other wavelengths, the vertical axis is in logarithmic scale, with M, C, B and A referring to the equivalent {\it GOES} flare categories. Associated photometric uncertainties are smaller than the size of diamond symbols and thus are not shown. The data in this and subsequent timeseries figures show prominent diurnal patterns: these are artifacts.\label{fig:lc_qs}}
\end{figure*}

For each case we generated light curves of a variety of wavelengths ranging from visible, UV, EUV, to soft X-ray. The data set and methods for light curve derivations are shown in the following subsections. For most of the generated light curves, we applied smoothing by filtering out the components with time scales shorter than 36 hr in the Fourier domain in order to minimize the effects of short-term variations such as solar flares and artificial noises. A mirror symmetric boundary condition was applied for the Fourier transforms. Figure \ref{fig:lc_qs} presents the light curves for the 20 days of QS conditions starting at 00:00 UT on 2019 November 27. The noise levels of these light curves are small, and result from instrumental variation rather than photon noise or readout noise, as discussed further in the following subsections.

\subsection{{\it SORCE}/TIM}

The TSI data sets were obtained by the Total Irradiance Monitor (TIM; \citealt{2005SoPh..230..129K}) aboard the {\it Solar Radiation and Climate Experiment} ({\it SORCE}; \citealt{2005SoPh..230....7R}). We downloaded the 6-hourly mean TSI data (Level 3) from the archive\footnote{\url{http://lasp.colorado.edu/home/sorce/data/tsi-data/}} and smoothed them by filtering out the $<$36-hr components. The top left panel of Figure \ref{fig:lc_qs} presents the TSI curve for the QS period. From the description of the TIM data, the noise level is assumed to be constant at 5 ppm.
%Note that in this panel, we show the noise level (constant at 5 ppm) as the typical error although the TSI may fluctuate in each 6-hr bin to a certain degree with the measurement uncertainty of $\sim 450$ ppm.

\subsection{{\it SDO}/HMI}

We used the full-disk continuum images taken by the Helioseismic and Magnetic Imager (HMI; \citealt{2012SoPh..275..207S,2012SoPh..275..229S}) on board the {\it Solar Dynamics Observatory} ({\it SDO}; \citealt{2012SoPh..275....3P}) to produce the visible (photospheric) light curves at 6173~{\AA}. The HMI line-of-sight (LOS) magnetograms were also used for measuring the variations of the total unsigned magnetic flux. We used the sets of synoptic magnetograms\footnote{\url{http://jsoc.stanford.edu/data/hmi/fits}} and continuum images, both of which are rebinned from the original $4096\times 4096$ pixels to $1024\times 1024$ by averaging the values in the $4\times 4$ pixel cell. The cadence and duration are 1 hr and 20 days (i.e., in total $\sim 480$ snapshots for each observable), respectively, whereas the pixel size and the entire field of view (FOV) are $2\farcs 0$ and $2048\arcsec\times 2048\arcsec$, respectively.

Because the sunspots and bright regions (faculae and plages) have only small areal coverages on the solar disk, with the maximum spot area being 240 MSH in this study, intensity reductions and enhancements in the light curves that are generated simply integrating the HMI continuum images over the entire solar disk are almost completely obscured by the diurnal, annual, and other variations caused by {\it SDO}'s orbital motion and the eccentricity of Earth's orbit, which are not easily removed \citep[e.g.,][]{2016SoPh..291.1887C}. Therefore, to minimize these effects, we normalized the disk-integrated intensity of each snapshot (irradiance) by that of the QS baseline image generated from the non-magnetized pixels in the same snapshot (see below).

For each HMI continuum image, we first masked out the magnetized ($\geq 5\ {\rm G}$) pixels using the simultaneous magnetogram, and averaged the non-magnetized pixels at each radial distance from the disk center in the azimuthal direction to create a one-dimensional QS profile as a function of radius. At the disk center, where the pixel numbers are small, the function was determined by fitting a cosine function. As a result, a QS baseline image was generated in such a way that each pixel of a given radial distance contains the intensity of the one-dimensional QS strip at the corresponding radial distance. Finally, we integrated the intensity in the original continuum image $I_{\rm HMI}$ over the disk, took the ratio to the counterpart of the QS baseline image $I_{\rm QS}$ as
\begin{eqnarray}
  F_{\rm vis}(t)=\int I_{\rm HMI}(t)\, dS\,\bigg/\,\int I_{\rm QS}(t)\, dS,
\end{eqnarray}
and repeated the above procedure for the set of $\sim 480$ frames to generate the HMI visible light curves.

In this light curve derivation, it is implicitly assumed that the continuum intensity of the QS pixels is constant during each 20-day observation window. Such an assumption may be justified by the work by \citet{2006SoPh..235..369W}, who found that the short-term (weeks to months) variation in TSI (or almost equivalently in visible) is mainly attributed to intensity variations of regions with enhanced magnetic fields of greater than 10 G. Also, as the ragged edges of the solar image due to the finite-sized pixels caused the noise in irradiance, we only used the pixels with radial distance of up to about $960\arcsec$ in the disk integration \citep{2006SoPh..235..369W}.

The resultant HMI visible continuum light curve for the QS period is shown in Figure \ref{fig:lc_qs} (top right). Here, the smoothed light curves are made by filtering out the $<$36-hr components after any observational gaps such as the spacecraft eclipse are filled by linearly interpolating the data. The observational error is estimated based on \citet{2016SoPh..291.1887C}. While the {\it SORCE} data are optimized for irradiance monitoring, we mainly use HMI visible light curves for analysis in the following sections because spatially resolved images allow us to identify which structure contributes to the light curves and the higher cadence enables better comparison with light curves of other wavelengths.

The total unsigned magnetic flux was calculated for each magnetogram by integrating the magnetic field strength $B_{\rm LOS}$ over the disk as
\begin{eqnarray}
  \Phi_{\rm LOS}(t)=\int |B_{\rm LOS}(t)|\, dS,
\end{eqnarray}
with similarly removing the pixels near the limb. We also corrected $B_{\rm LOS}$ for the LOS projection by dividing it by $\cos{\theta}$, where $\theta$ is the viewing angle from the disk center, and obtained the (approximately) radial magnetic flux:
\begin{eqnarray}
  \Phi_{\rm rad}(t)=\int \frac{|B_{\rm LOS}(t)|}{\cos{\theta}}\, dS.
\end{eqnarray}
The noise estimation is based on \citet{2001ApJ...555..448H}. As the HMI total flux curves are known to present oscillations of 24-hr and other time scales \citep[see, e.g.,][]{2014ApJ...794...19T,2017SoPh..292...48A}, they are smoothed by removing the $<$36-hr components. The original and smoothed total flux curves are presented in Figure \ref{fig:lc_qs} (second row).

\subsection{{\it SDO}/AIA}

\begin{figure}
\begin{center}
\includegraphics[width=0.5\textwidth]{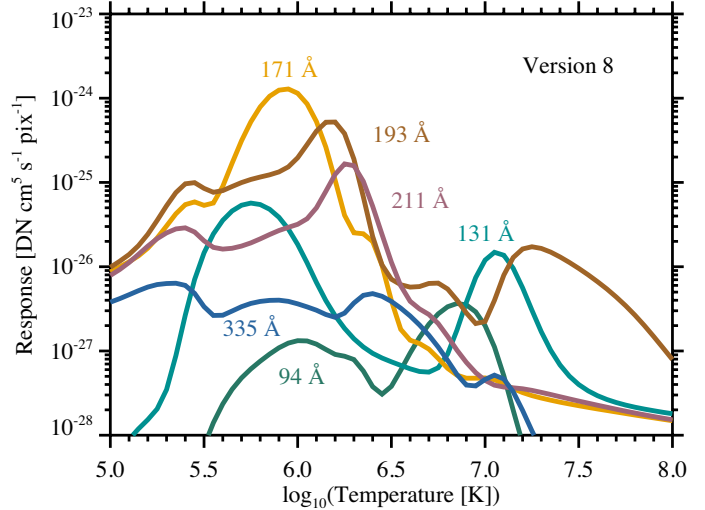}
\end{center}
\caption{Temperature response functions for the six EUV channels of {\it SDO}/AIA. Shown here is the version 8 issued on 2017 December 11, i.e., applicable to all AIA data sets used in this study. Here 304~{\AA} is not shown because the \ion{He}{2} line at 304~{\AA} is not well modeled under the assumptions of statistical equilibrium \citep{2014ApJ...784...30G} and therefore is typically not used for DEM inversions \citep{2005ApJS..157..147W,2012SoPh..275...41B}.\label{fig:aiarf}}
\end{figure}

UV and EUV images are captured by {\it SDO}'s Atmospheric Imaging Assembly (AIA; \citealt{2012SoPh..275...17L}). We used the AIA synoptic data\footnote{\url{http://jsoc.stanford.edu/data/aia/synoptic}}, which are generated by processing the original $4096\times 4096$-pixel Level 1 data to Level 1.5 and reducing them to $1024\times 1024$ pixels through $4\times 4$-pixel averaging. The data set is composed of 1600~{\AA}, 1700~{\AA}, 304~{\AA}, 171~{\AA}, 193~{\AA}, 211~{\AA}, 335~{\AA}, 131~{\AA}, and 94~{\AA} channel images, each having the pixel size of $2\farcs 4$, FOV of $2458\arcsec\times 2458\arcsec$, cadence of 1 hr, and duration of 20 days (i.e., about 480 frames each depending on the {\it SDO} eclipse). The characteristic temperature and representative spectral lines of each AIA bandpass are shown in Table \ref{tab:aia} (\citealt{2012SoPh..275...17L}; see also \citealt{2010A&A...521A..21O} for thorough line lists). However, one should be reminded that each band has wide temperature sensitivity, which is characterized by the temperature response function in Figure \ref{fig:aiarf}.

\begin{deluxetable*}{cccc}
%\tablenum{1}
\tablecaption{Primary Ions and Characteristic Temperatures of the AIA Channels\label{tab:aia}}
\tablewidth{0pt}
\tablehead{
\colhead{Channel [{\AA}]} & \colhead{Primary ion(s)} & \colhead{Atmosphere} & \colhead{$\log(T\,{\rm [K]})$}
}
%\decimalcolnumbers
\startdata
1600 & continuum ($+$\ion{C}{4}) & upper photosphere, transition region & 3.7 (5.0)\\
1700 & continuum & photosphere & 3.7\\
304 & \ion{He}{2} & chromosphere, transition region & 4.7\\
171 & \ion{Fe}{4} & quiet corona, upper transition region & 5.8\\
193 & \ion{Fe}{12}, \ion{Fe}{24} & corona, hot flare plasma & 6.2, 7.3\\
211 & \ion{Fe}{14} & active-region corona & 6.3\\
335 & \ion{Fe}{16} & active-region corona & 6.4\\
131 & \ion{Fe}{8}, \ion{Fe}{21} & transition region, flaring corona & 5.6, 7.0\\
94 & \ion{Fe}{18} & flaring corona & 6.8\\
\enddata
\tablecomments{Table based on \citet{2012SoPh..275...17L}.}
\end{deluxetable*}

Because it is difficult to generate AIA QS baseline images as we did for the HMI light curves, we assumed that the detector degradation was negligibly small over each observation window (i.e., 20 days) and, after normalizing them by the exposure time and the Sun-{\it SDO} distance, simply integrated the intensities in each frame:
\begin{eqnarray}
  F_{\rm AIA}(t)=\int I_{\rm AIA}(t)\, dS.
\end{eqnarray}
%The Doppler velocity (relativistic) correction was not applied here as its effect is negligibly small for the fluctuations in AIA light curves\footnote{Because the incoming photons have more energy by a factor $(1-\beta)$ and the instruments collect photons at a rate larger by a factor $(1-\beta)$, the irradiance varies in proportion to $(1-\beta)^{2}\approx 1-2\beta$, where $\beta=V_{\rm D}/c$, $V_{\rm D}$ the spacecraft velocity relative to the Sun, and $c$ the speed of light. As the orbital velocity of {\it SDO} is about $3.5\ {\rm km\ s}^{-1}$ \citep{2016SoPh..291.1887C}, the Doppler correction factor is only up to about 20 ppm, which is much smaller than the fluctuations of AIA light curves shown in Figure \ref{fig:lc_qs}.}.
The resulting light curves were smoothed by suppressing the $<$36-hr components after any gaps were filled.
%The noise levels were estimated in the four corners of each AIA image far above the limb as the ``roughness'' of the image, which was the standard deviation of the residual of each pixel from the azimuthally-averaged intensity at the corresponding radial distance.
Noise levels were estimated by using the AIA SolarSoftWare routine {\tt aia\_bp\_estimate\_error.pro} and were much less than 0.1\% for all channels.

The resultant unsmoothed and smoothed AIA light curves for the QS period are presented in Figure \ref{fig:lc_qs}.
Variation amplitudes of the diurnal oscillations in some of the EUV channels (335~{\AA} and 94~{\AA}) are relatively large because the data numbers (DNs) in these channels are small. Typical DNs of these passbands are of order unity for the time period considered.
%So the absolute diurnal variations are only fractions of a DN.
(In these cases, the $\sim 0.1$~K change in the CCD temperature affects the DNs at the sub-DN level, in turn causing the disk-integrated DN to vary by a few percent. Such diurnal variations are attenuated in the following analysis by the 36-hr smoothing.)

\subsection{{\it Hinode}/XRT}

The X-Ray Telescope (XRT; \citealt{2007SoPh..243...63G}) aboard the {\it Hinode} spacecraft \citep{2007SoPh..243....3K} routinely obtains the full-disk synoptic soft X-ray images \citep{2016SoPh..291..317T}\footnote{\url{http://solar.physics.montana.edu/HINODE/XRT/SCIA/latest_month.html}}. Currently the synoptic data are taken twice a day around 6 and 18 UT although often shifted or skipped due to other conditions (XRT bakeout, {\it Hinode} eclipse, etc.).

From the downloaded XRT data, we left out the images which were taken outside of the $\pm 15$-minute window centered at 6 or 18 UT, or whose pointings are deviated by more than $5\arcsec$. For reproducing the disk-integrated light curves, we used Al\_mesh and Al\_poly filter images, both peaked around $\log{T}=6.9$ with the highest responses to the temperature of all available XRT filters \citep{2011SoPh..269..169N}. Each $1024\times 1024$-pixel image has a spatial sampling of $2\arcsec$ (i.e. $2048\arcsec\times 2048\arcsec$ FOV). The XRT irradiance is obtained by simply integrating the pixels over each frame:
\begin{eqnarray}
  F_{\rm XRT}(t)=\int I_{\rm XRT}(t)\, dS.
\end{eqnarray}
Because the cadence is 12 hr or longer (and is not constant), we did not apply the 36-hr smoothing (see Figure \ref{fig:lc_qs}).
%The noise estimation is similar to those for the AIA data 

\subsection{{\it GOES}/XRS}

In addition, we plotted the {\it GOES}/XRS (1--8~{\AA} band) light curves. We used the ``science quality'' Level 2 data from {\it GOES}-16 satellite (1-min average)\footnote{\url{https://www.ngdc.noaa.gov/stp/satellite/goes-r.html}}. In Figure \ref{fig:lc_qs}, the {\it GOES} flux is steady around $4.2\times 10^{-9}\ {\rm W\ m^{-2}}$.

\section{Transit Light Curves}\label{sec:results}

%In this section, we present the light curves for the transiting isolated sunspot (Sections \ref{subsec:sunspot}), plage (Section \ref{subsec:plage}), and emerging flux (\ref{subsec:ef}).

\subsection{Sunspot: AR 12699}\label{subsec:sunspot}

\begin{figure*}
\begin{center}
\includegraphics[width=0.9\textwidth]{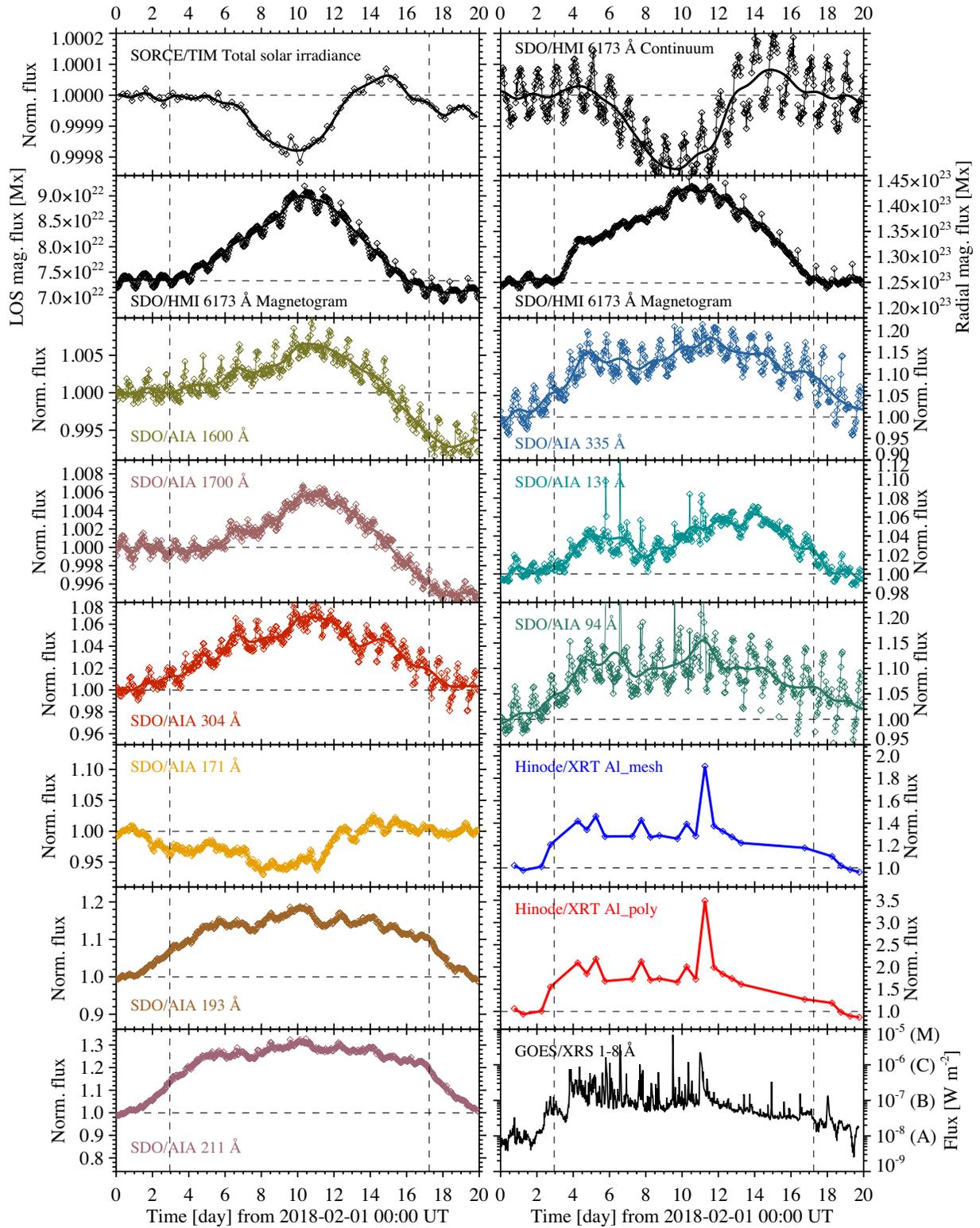}
\end{center}
\caption{Same as Figure \ref{fig:lc_qs} but for the transiting sunspot (AR 12699) starting at 00:00 UT on 2018 February 1. Light curves are normalized by the mean value obtained by averaging the smoothed lines over the first 24 hr (represented by the horizontal dashed lines), while the times when the AR appears at the eastern limb and disappears at the western limb are indicated by the vertical dashed lines. Note that the vertical scales are consistent with those in Figure \ref{fig:lc_qs}, highlighting that the amplitudes of rotational modulation presented in this figure are much larger than the QS fluctuation levels. \label{fig:lc_spot1}}
\end{figure*}

\begin{figure*}
%\begin{interactive}{animation}{12.mp4}
\begin{center}
\includegraphics[width=0.95\textwidth]{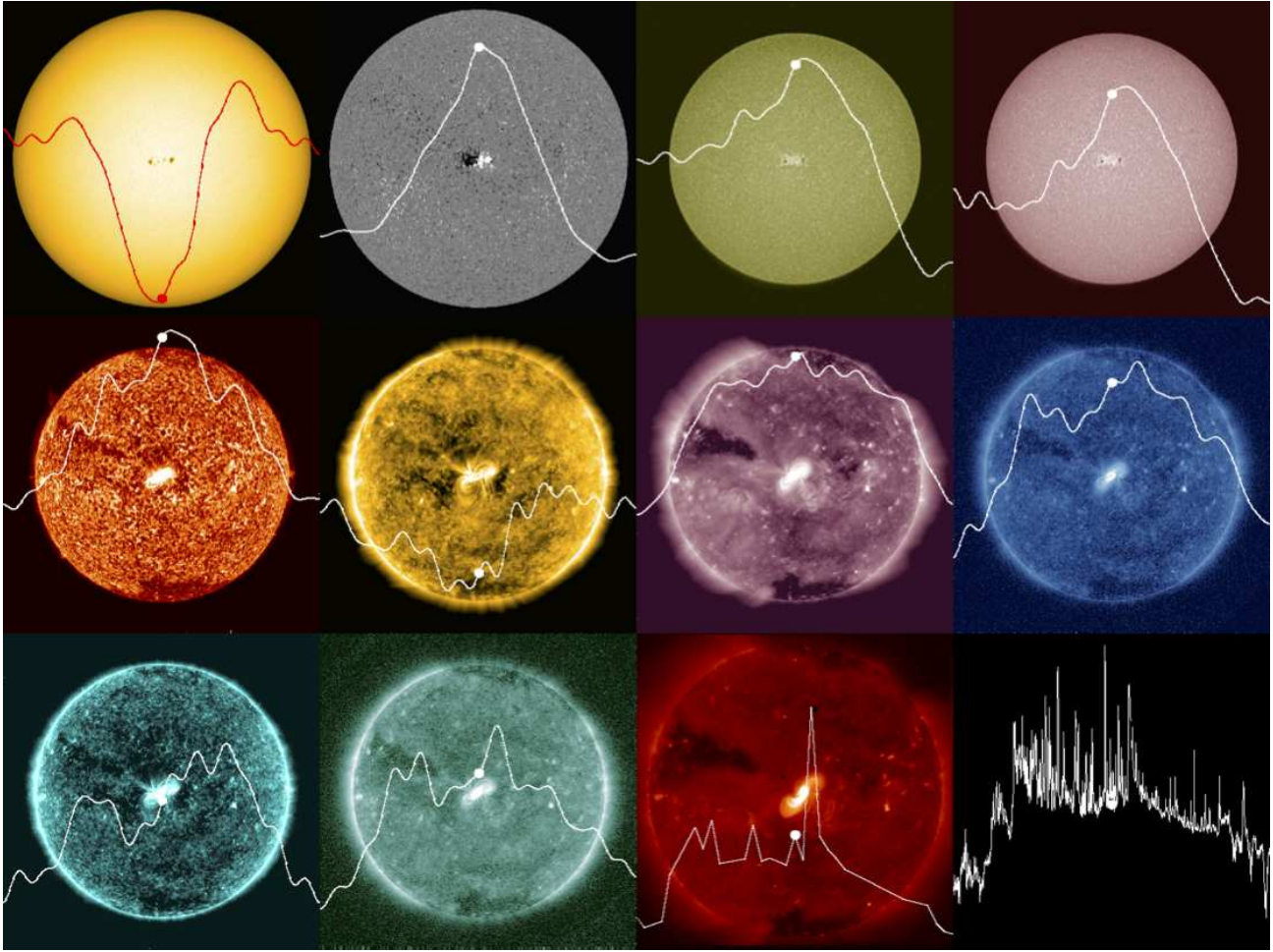}
\end{center}
%\end{interactive}
\caption{Mosaic of the full-disk images for the transiting sunspot (AR 12699) with the corresponding light curves overlaid. Shown here are the still images at 00:00 UT on 2018 February 11. From top left to bottom right, {\it SDO}/HMI continuum, HMI magnetogram, {\it SDO}/AIA 1600~{\AA}, 1700~{\AA}, 304~{\AA}, 171~{\AA}, 211~{\AA}, 335~{\AA}, 131~{\AA}, 94~{\AA}, {\it Hinode}/XRT Al\_poly, and {\it GOES}/XRS (no image). The dynamic ranges of the light curves are consistent with those in Figure \ref{fig:lc_spot1}. An animation of this figure is available. It covers 20 days of observing beginning at 00:00 UT, 2018 February 1 and the video duration is 8 s.\label{fig:tile_spot1}}
\end{figure*}

We first analyze the isolated sunspot group AR 12699, whose light curves are summarized as Figure \ref{fig:lc_spot1}. Figure \ref{fig:tile_spot1} shows the mosaic of the corresponding images with the light curves (see the accompanying video for the temporal evolution). This region rotated over the eastern limb and appeared on the visible solar disk on 2018 February 3. It passed the central meridian just at the disk center on February 11 with the maximum (reported) spot area of 240 MSH. A number of solar flares including four C-class events were produced before it left the visible disk on February 17. Throughout this period, there were several minor regions seen on the disk, but all of them were short-lived, decaying, spotless plage regions.

The light curves in Figure \ref{fig:lc_spot1} are normalized by the mean values of the smoothed curves during the first 24 hr, i.e., when the solar disk remained quiescent. The timings when the AR appeared on the eastern limb and disappears at the western limb are overplotted, which are determined by visually inspecting the photospheric images.

From the TSI and visible continuum (HMI 6173~{\AA}) light curves, it is evident that the spot causes a remarkable reduction in irradiance (``dip'') around the central meridian transit with two enhancements (``shoulders'') when the spot is near the east and west limbs. The amplitudes of the dips are 170 and 240 ppm for the TSI and visible, respectively, and their maximum enhancements are 70 and 80 ppm, respectively.
%These values are much greater than the 3$\sigma$ or 5$\sigma$ levels of the QS variations in Figure \ref{fig:lc_qs}.
These values are much greater than the QS variations in Figure \ref{fig:lc_qs}.
The reduction in the visible irradiance of 240 ppm is quite reasonable considering that the maximum spot size is 240 MSH at the disk center. These features are well in line with the previous TSI studies introduced in Section \ref{sec:intro}.

The total unsigned LOS magnetic flux in the photosphere (HMI magnetogram) attains its peak value at $9.0\times 10^{22}\ {\rm Mx}$ on February 11, marking the enhancement of 22.8\% from the reference QS level. The radial magnetic flux curve shows more flattened profile than the LOS flux. The peak value is $1.4\times 10^{23}\ {\rm Mx}$, which is enhanced by 14.7\% from its reference.

All AIA UV and EUV light curves, except for 171~{\AA}, show mountain-shaped or flat-top profiles, and the peak magnitude generally increases with the characteristic temperature (see Table \ref{tab:aia} and Figure \ref{fig:aiarf} for the corresponding characteristic temperatures). The reduction of the 171~{\AA} irradiance is discussed further in Section \ref{subsec:171}. One may notice that the unsmoothed high-temperature light curves from the 131~{\AA} and 94~{\AA} channels of AIA, which are sensitive to flaring plasmas, exhibit distinct spikes with magnitudes far exceeding the diurnal fluctuations. On closer inspection, it is clear that these spikes appear concurrently with the {\it GOES} X-ray flares. Some of the long-duration ($>$1 hr) events are temporally resolved.

The general trend of higher-temperature light curves having larger peak magnitudes also applies to the soft X-ray curves. The background levels are enhanced by 30\% (XRT Al\_mesh) to more than one order of magnitude ({\it GOES}) from the QS level. Although the cadence is coarser than the other light curves, parts of the XRT synoptic observations coincide with the X-ray flares and show drastic enhancements in irradiance (e.g., the C1.5-class event SOL2018-02-12T00:15).

The light curves that are sensitive to the transition-region and coronal plasmas (i.e., EUV channels of 304~{\AA} or higher and soft X-ray observables) start to become enhanced about two days before the appearance of the AR in the photosphere (indicated by the vertical dashed lines). Together with its counterpart when the AR rotates beyond the western limb, these time lags reflect the extended coronal magnetic structures stretching above the solar limbs. Geometrically, the two-day advance appearance could be accounted by a loop system of approximate height $\sim$\, 40 Mm, consistent with what is seen in the corresponding AIA images (see movie of Figure \ref{fig:tile_spot1}).

%We discuss the observed time lags in Section \ref{subsec:lag}.

\subsection{Plage: Return of AR 12713}\label{subsec:plage}

\begin{figure*}
\begin{center}
\includegraphics[width=0.9\textwidth]{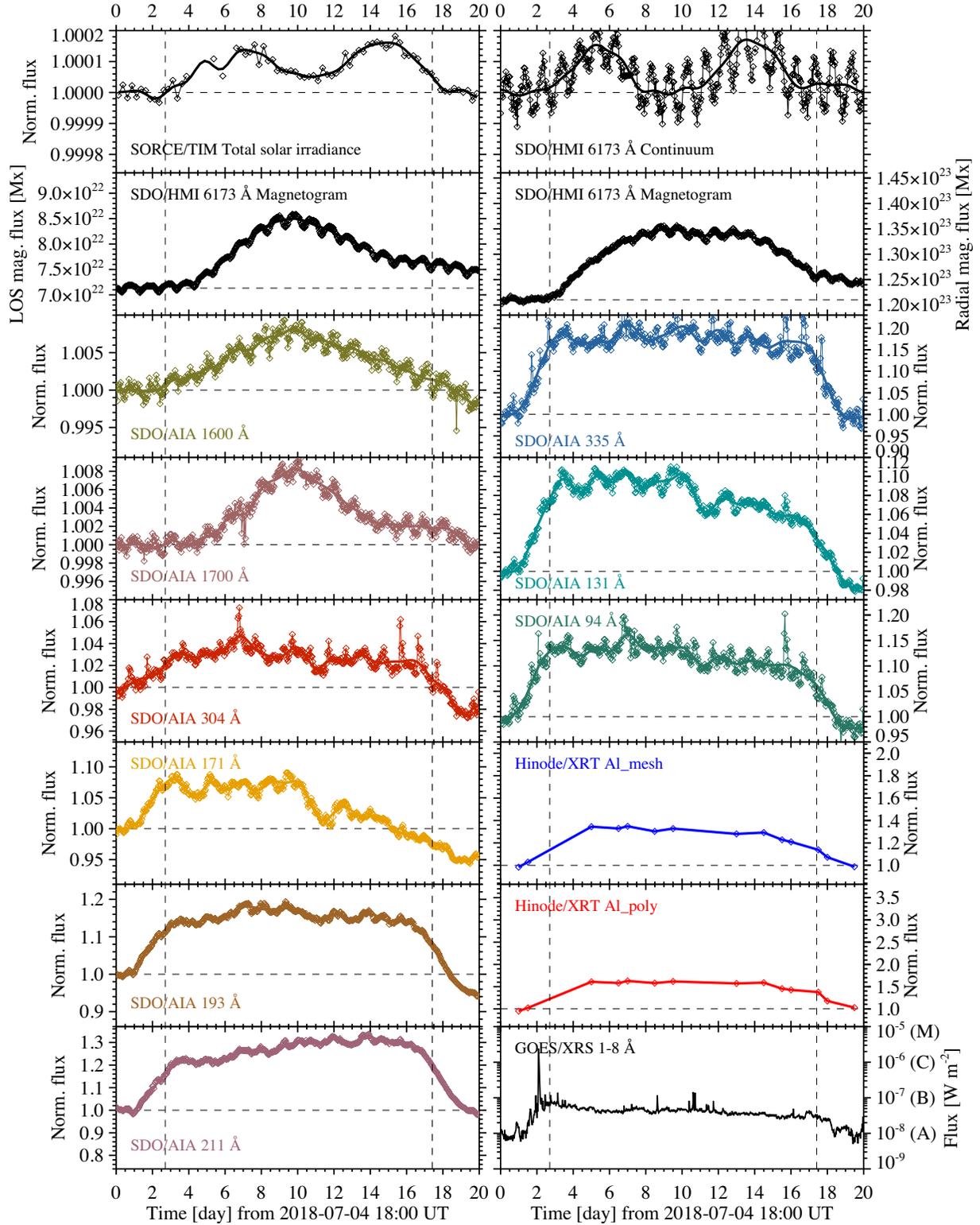}
\end{center}
\caption{Same as Figure \ref{fig:lc_spot1} but for the transiting plage (return of AR 12713) starting at 18:00 UT on 2018 July 4.\label{fig:lc_plage1}}
\end{figure*}

\begin{figure*}
%\begin{interactive}{animation}{15.mp4}
\begin{center}
\includegraphics[width=0.95\textwidth]{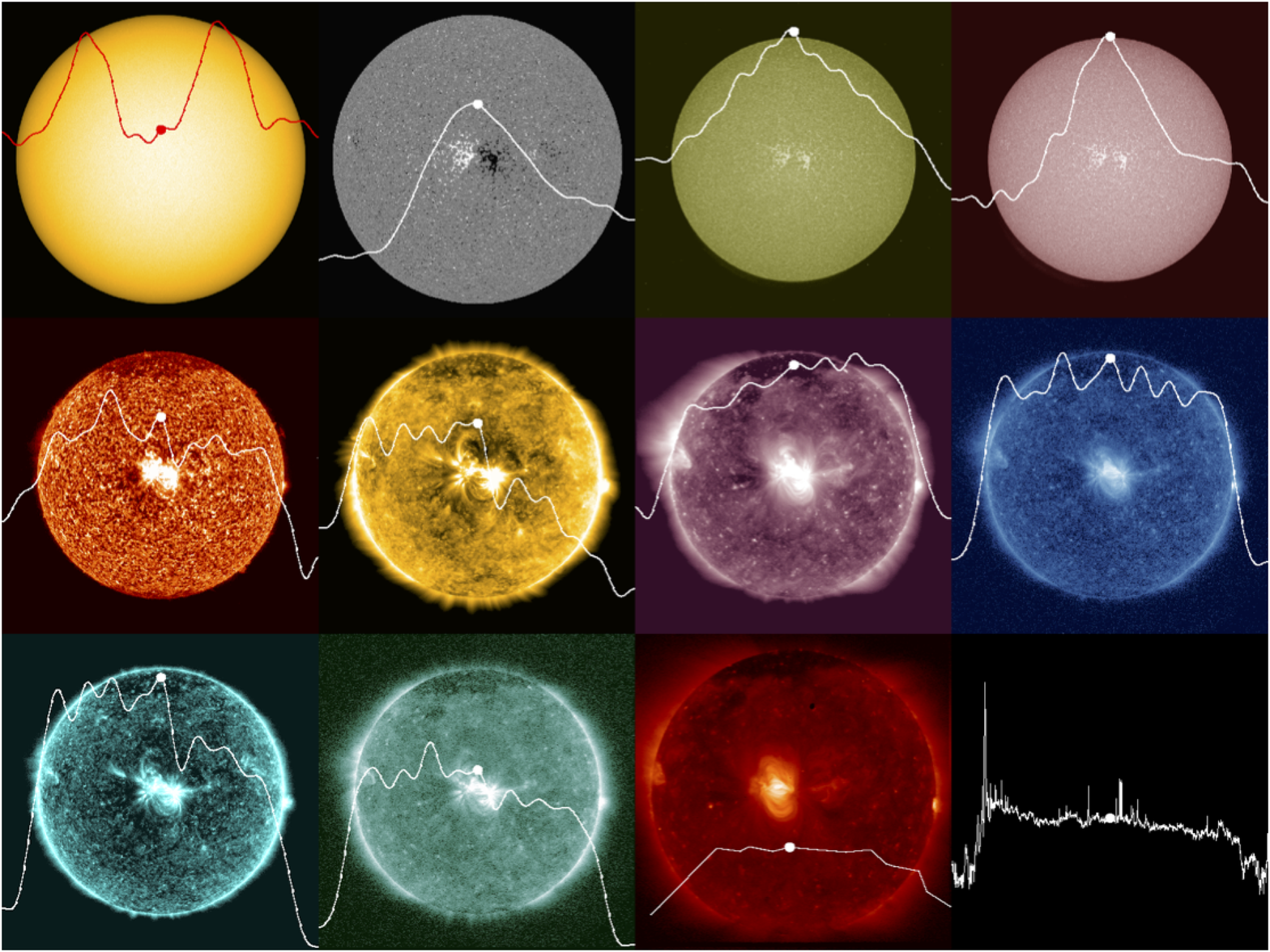}
\end{center}
%\end{interactive}
\caption{Same as Figure \ref{fig:tile_spot1} but for the transiting plage (return of AR 12713). Shown here are the still images at 18:00 UT on 2018 July 14. An animation of this figure is available. It covers 20 days of observing beginning at 18:00 UT, 2018 July 4 and the video duration is 8 s.\label{fig:tile_plage1}}
\end{figure*}

The second case is the spotless plage region in July 2018, which is the decay product of AR 12713 from the previous rotation. This plage region rotated into the visible hemisphere on July 7 and disappeared at the western limb on July 22. Its light curves are shown in Figures \ref{fig:lc_plage1} and \ref{fig:tile_plage1}.

The visible continuum (HMI 6173~{\AA}) and TSI curves have two ``shoulders'' akin to the sunspot case. However, these curves do not possess ``dips'' when the plage is at the central meridian, and the TSI is even increased from the reference QS level. The amplitude of the ``shoulders'' are larger than those of the sunspot case, with the maximum enhancement in the visible continuum being 170 ppm. The total LOS and radial magnetic fluxes are $8.5\times 10^{22}\ {\rm Mx}$ (increased by 19.2\% from the reference level) and $1.3\times 10^{23}\ {\rm Mx}$ (increased by 11.4\%).

Most of the AIA UV and EUV curves show trends similar to those of the spot case. The light curve from the 171~{\AA} channel, in contrast to its anti-phased profile for the transiting sunspot case, now shows a brightening especially in the first half of the plage transit, in line with the trends of the other EUV curves. Unfortunately there are not many XRT observations during this period, partly due to an XRT bakeout. Therefore, the pointing condition of $\leq 5\arcsec$ is relaxed to keep enough data points. Slightly before this plage rotates onto the visible disk, a C1.6-class flare occurs (SOL2018-07-06T19:41) as observed in the AIA 335~{\AA} and 94~{\AA} channels. Similarly to the transiting sunspot case, starting and ending of the transition-region and coronal irradiances of this event reveal time lags from the photospheric and chromospheric counterparts (vertical dashed lines).

\subsection{Emerging Flux: AR 12733}\label{subsec:ef}

\begin{figure*}
\begin{center}
\includegraphics[width=0.9\textwidth]{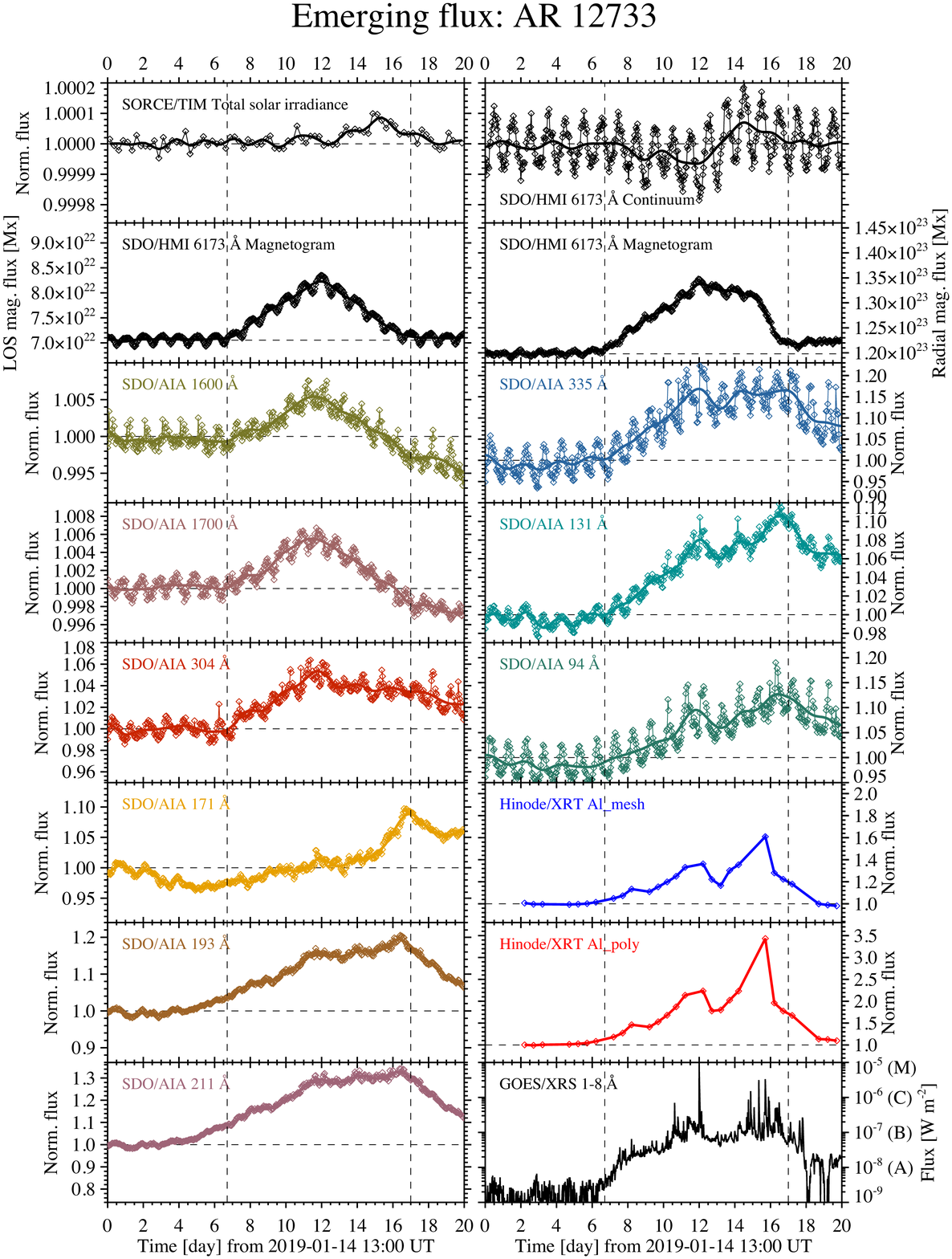}
\end{center}
\caption{Same as Figure \ref{fig:lc_spot1} but for the transiting emerging flux (AR 12733) starting at 13:00 UT on 2019 January 14.\label{fig:lc_ef1}}
\end{figure*}

\begin{figure*}
%\begin{interactive}{animation}{03.mp4}
\begin{center}
\includegraphics[width=0.95\textwidth]{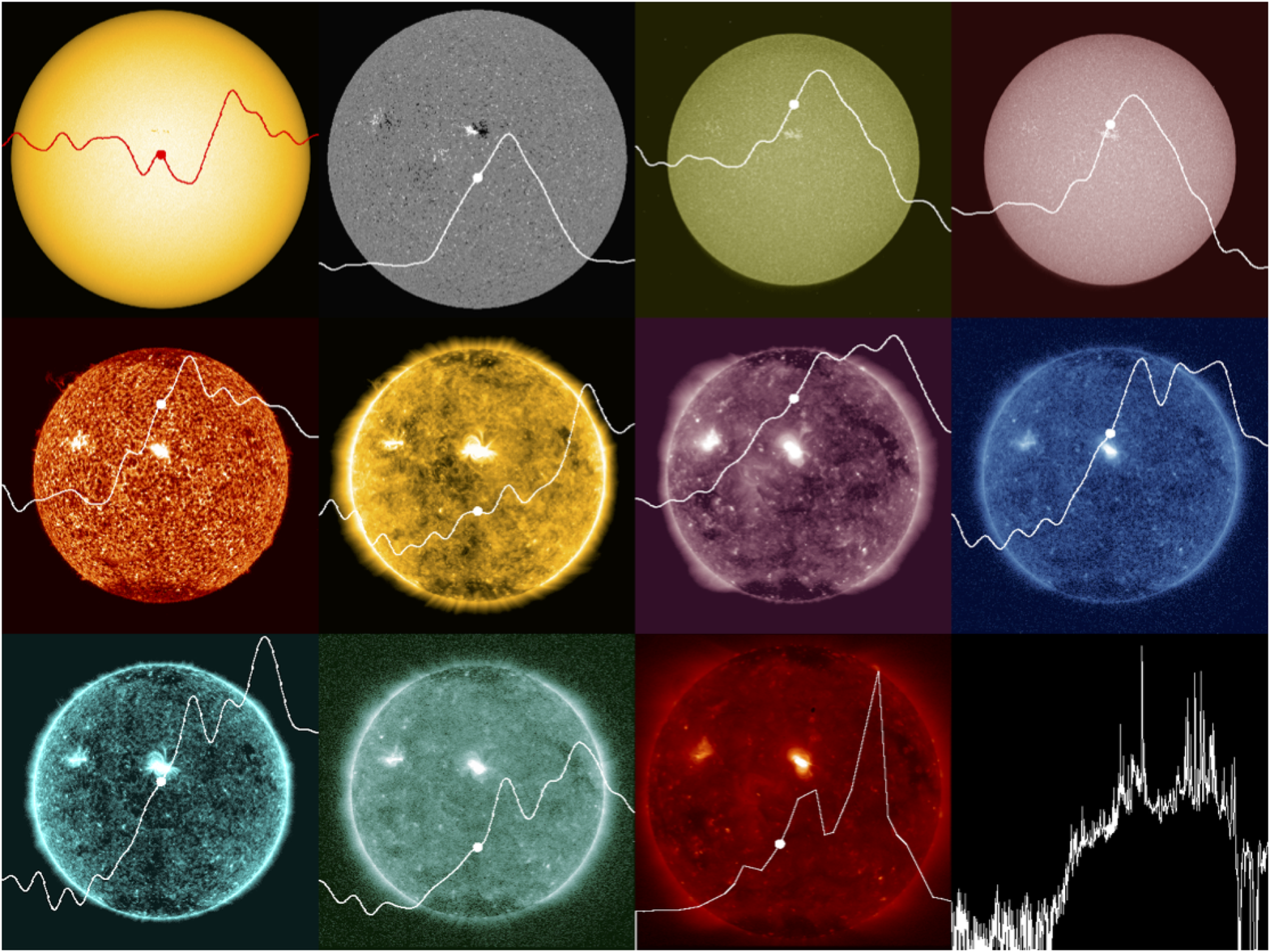}
\end{center}
%\end{interactive}
\caption{Same as Figure \ref{fig:tile_spot1} but for the transiting emerging flux (AR 12733). Shown here are the still images at 13:00 UT on 2019 January 24. An animation of this figure is available. It covers 20 days of observing beginning at 13:00 UT, 2019 January 14 and the video duration is 8 s.\label{fig:tile_ef1}}
\end{figure*}

We also analyze light curves corresponding to the emergence of AR 12733, which emerged in the eastern hemisphere on 2019 January 21, crossed the central meridian on January 24, and left the western limb on January 31 (Figures \ref{fig:lc_ef1} and \ref{fig:tile_ef1}). Now the visible continuum, TSI, and magnetic flux curves are all quite asymmetric about the central meridian. The visible continuum shows a slight reduction (``dip'') and enhancement (``shoulder''), both amounting to 70 ppm. The maximum spot size is recorded to be 90 MSH on the day of west limb transit (January 31). The total LOS and radial magnetic fluxes are $8.2\times 10^{22}\ {\rm Mx}$ (increased by 16.7\% from the reference QS level) and $1.3\times 10^{23}\ {\rm Mx}$ (11.7\%), respectively.

The most noticeable feature of the UV, EUV, and soft X-ray curves from the sunspot and plage cases is their asymmetry. Because the AR continued to develop, the EUV curves are peaked on the day when it passes the western limb (January 31: vertical dashed line). However, the slow decay of these curves after that day is likely affected by another spotless plage region that is appearing at the eastern limb. Several X-ray flares, especially a C5.0-class event (SOL2019-01-26T13:12), are seen in AIA 335 and 131~{\AA} light curves.

\section{Detailed Analysis on the Light Curves}\label{sec:detailed}

\subsection{Correlations between UV/EUV Irradiances and Photospheric Magnetic Flux}

\begin{deluxetable}{cCCC}
%\tablenum{1}
\tablecaption{Correlations between UV/EUV Irradiance and Photospheric Magnetic Flux\label{tab:cc}}
\tablewidth{0pt}
\tablehead{
\colhead{Channel [{\AA}]} & \colhead{Sunspot} & \colhead{Plage} & \colhead{Emerging flux}
}
%\decimalcolnumbers
\startdata
$F_{\rm AIA}$ vs $\Phi_{\rm LOS}$ & & &\\
1600 & {\bf 0.89} & {\bf 0.92} & {\bf 0.88}\\
1700 & {\bf 0.92} & {\bf 0.97} & {\bf 0.92}\\
304 & {\bf 0.93} & 0.44 & 0.77\\
171 & -0.63 & 0.26 & -0.05\\
193 & 0.79 & 0.59 & 0.60\\
211 & 0.76 & 0.78 & 0.57\\
335 & 0.78 & 0.57 & 0.63\\
131 & 0.64 & 0.48 & 0.45\\
94 & 0.78 & 0.45 & 0.39\\
%\hline
$F_{\rm AIA}$ vs $\Phi_{\rm rad}$ & & &\\
1600 & {\bf 0.85} & {\bf 0.93} & 0.76\\
1700 & {\bf 0.88} & {\bf 0.86} & 0.79\\
304 & {\bf 0.97} & 0.62 & {\bf 0.86}\\
171 & -0.52 & 0.34 & -0.12\\
193 & {\bf 0.89} & 0.74 & 0.76\\
211 & {\bf 0.87} & {\bf 0.84} & 0.79\\
335 & {\bf 0.90} & 0.71 & 0.78\\
131 & 0.79 & 0.63 & 0.62\\
94 & {\bf 0.88} & 0.60 & 0.58\\
\enddata
\tablecomments{The values shown here are the linear Pearson correlation coefficients between the smoothed AIA UV/EUV light curves, $F_{\rm AIA}$, and the smoothed total unsigned LOS and radial magnetic fluxes of HMI, $\Phi_{\rm LOS}$ and $\Phi_{\rm rad}$. Values with higher correlations ($>0.8$) are shown in bold.}
\end{deluxetable}

In order to investigate the relationships between the UV/EUV irradiances and the total magnetic flux in the photosphere, we compute the linear Pearson correlation coefficients between the smoothed AIA light curves for all nine bands ($F_{\rm AIA}$) and the smoothed total unsigned magnetic fluxes for both LOS ($\Phi_{\rm LOS}$) and radial ($\Phi_{\rm rad}$) components. The result is summarized in Table \ref{tab:cc}.

The strongest correlations with correlation coefficients $\gtrsim 0.8$ are always found with regard to the channels with low characteristic temperatures, i.e., 1600~{\AA} and 1700~{\AA}. According to \citet{2019ApJ...870..114S}, the strongest contributors to the plage emissions in AIA 1600~{\AA} and 1700~{\AA} are the continuum emissions that are formed near the temperature minimum (rather than the emission lines like \ion{C}{4}). The continuum contributions amount to 67\% and 87\% in 1600~{\AA} and 1700~{\AA}, respectively. Therefore, it is expected that these irradiances are sensitive to the evolution of apparent area of the magnetized regions, which may lead to the better correlations with $\Phi_{\rm LOS}$ than $\Phi_{\rm rad}$. However, in these AIA bandpasses, sunspots are dark and, therefore, should have a weakening effect to the correlations. This issue is further discussed in Section \ref{subsec:spotplage}.

In general, the correlations of the remaining channels, i.e., 304~{\AA} through 94~{\AA}, become worse with the characteristic temperature. For each individual channel, the correlations are better with $\Phi_{\rm rad}$ than $\Phi_{\rm LOS}$, which is opposed to the trend of the low temperature channels, and some of the correlation coefficients with $\Phi_{\rm rad}$ are greater than 0.8. These high temperature channels are sensitive to the coronal plasmas, which are optically thin. Therefore, once the target region is visible to the observer, only the intrinsic evolution such as the growth of magnetic flux matters, not the viewing angle. This may be the reason for better correlations with $\Phi_{\rm rad}$.

The exceptional case is 171~{\AA}, which presents almost zero to even negative correlations against the magnetic flux. The cause of this anti-phased evolution of 171~{\AA} is discussed in Section \ref{subsec:171}.

\subsection{Darkening and Brightening in Visible and UV Light Curves}\label{subsec:spotplage}

\begin{figure}
\begin{center}
\includegraphics[width=0.5\textwidth]{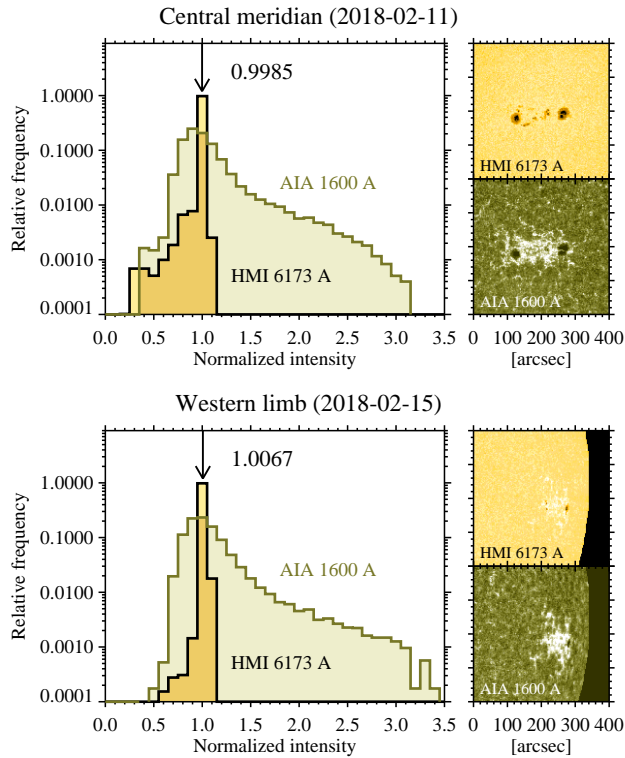}
\end{center}
\caption{Histograms of the normalized intensities of HMI 6173~{\AA} and AIA 1600~{\AA} for the two periods, 2018 February 11, when the sunspot AR 12699 is at the central meridian, and 2018 February 15, when the spot is near the limb (the visible light curve is at its peak). The $400\arcsec\times 400\arcsec$ areas are analyzed, which are shown on the right, but the off-limb region is not used.\label{fig:spotplage}}
\end{figure}

The visible continuum (HMI 6173~{\AA}) light curve of the transiting sunspot case (Figures \ref{fig:lc_spot1} and \ref{fig:tile_spot1}) shows the reduction in irradiance (``dip'') when the spot is around the central meridian and the brightenings near the limbs (``shoulders''). This indicates that, around the central meridian, the darkening effect of the sunspot umbra and penumbra dominates the brightening effect of the faculae around the spots, whereas the faculae have a prevailing effect near the limbs.

Figure \ref{fig:spotplage} compares the histograms of the normalized intensities in the two cutout HMI continuum images ($400\arcsec\times 400\arcsec$) at the central meridian (top) and at the western limb (bottom: when the visible light curve is at its maximum). In the present analysis, the limb darkening effect is removed. The histogram for the limb (bottom) reveals that there are many more bright pixels in which the normalized intensity exceeds unity ($I_{\rm HMI}/\langle I_{\rm HMI}\rangle >1$) compared with the central meridian case (top), as is evident from the accompanying HMI images. Even though the umbrae and penumbrae are still present at western limb, the extended faculae have a dominant contribution. As a result, the average intensity within the $400\arcsec\times 400\arcsec$ window is 1.0067 $(>1)$ for the limb, leading to the formation of the ``shoulders'', while it is 0.9985 $(<1)$ for the central meridian, i.e., the ``dip'' is formed.

Although the UV bands, such as AIA 1600~{\AA} and 1700~{\AA}, exhibit no ``dips'' when the spot is at central meridian, the similar competition for dominance between the spot and faculae should also exist because these channels are sensitive not only to the chromosphere but also to the photospheric continuum (Table \ref{tab:aia}). The histograms for 1600~{\AA} in Figure \ref{fig:spotplage} show that, at the central meridian, although the dark pixels ($I_{\rm 1600}/\langle I_{1600}\rangle<1$) do exist, there are a lot more bright pixels ($I_{\rm 1600}/\langle I_{1600}\rangle>1$). This overwhelms the darkening effect of the spots and produces the mountain-shaped profiles for the 1600~{\AA} irradiance, which is correlated well with the LOS magnetic flux.

\subsection{Anti-phased Variations of 171~{\AA} Irradiance}\label{subsec:171}

\begin{figure*}
\begin{center}
\includegraphics[width=0.85\textwidth]{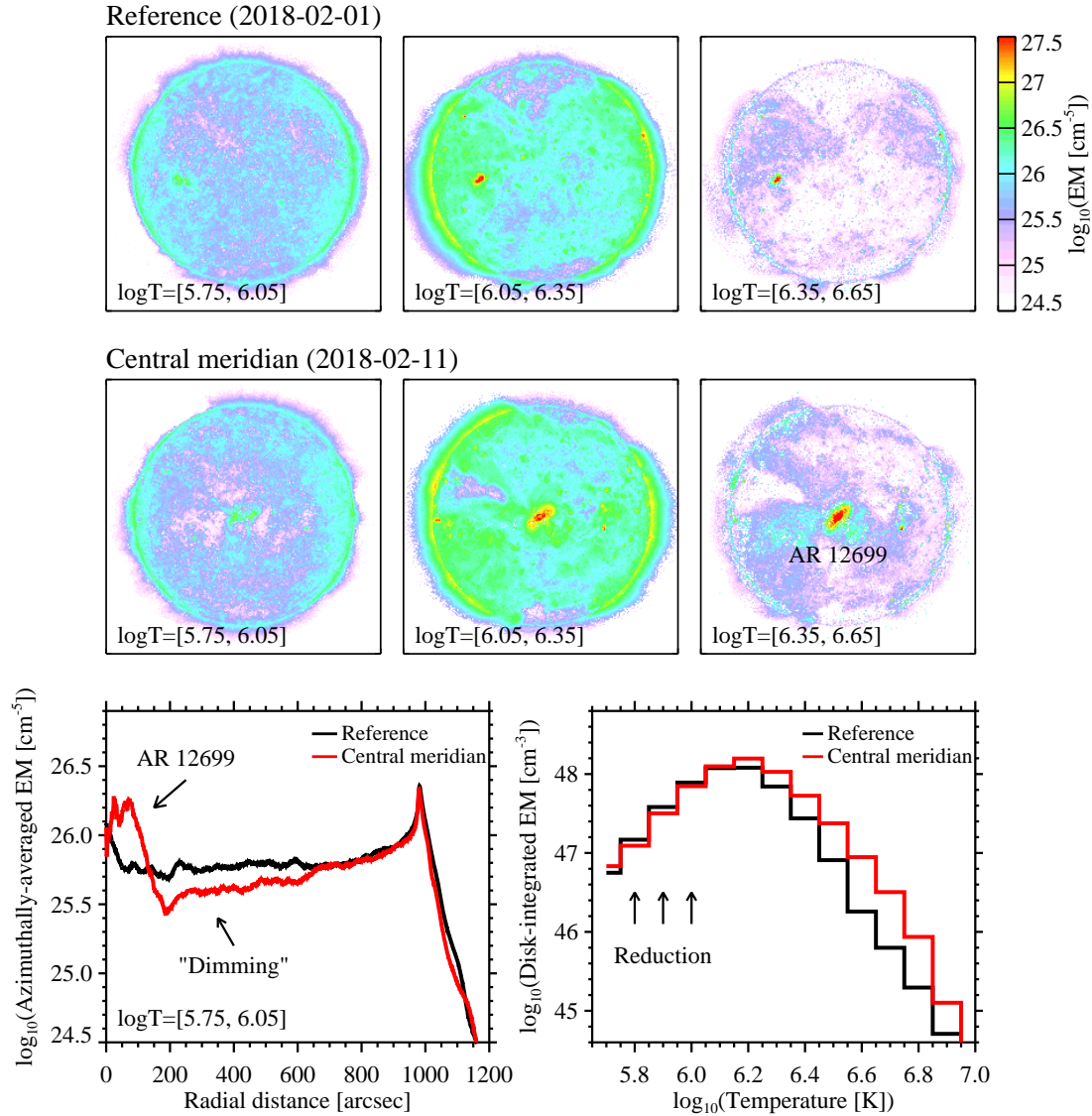}
\end{center}
\caption{DEM maps of the transiting sunspot AR 12699. The panels shown are the snapshots at 2018 February 01, when the target AR is behind the limb (reference: top), and 2018 February 11, when the AR is at the central meridian (middle). The color indicates the total EM within the $\log{T}$ bin indicated in each panel. The bottom left panel presents the azimuthally-averaged EMs for $\log{T}=[5.75, 6.05]$ as a function of the radial distance from the disk center, whereas the bottom right depicts the EMs integrated over the entire AIA FOV. The locations of AR 12699, the EM ``dimming'', and the bins where EMs are reduced from the reference are indicated with arrows. For this particular study, the original $4096\times 4096$-pixel full-resolution AIA images are used. Note that the pixels where the DEM code does not find the solutions due to insufficient DNs (about 30\% each), which are mostly located far above the limbs, and where unreasonably high temperature components are obtained (about 30\% each) are removed before the integration.\label{fig:dem}}
\end{figure*}

The outlier amongst the UV, EUV, and soft X-ray light curves is the 171~{\AA} channel of AIA, which shows almost zero to even anti-phased variations (Table \ref{tab:cc}). Namely, although ARs are observed in emission in 171~{\AA}, the disk-integrated irradiance can be reduced from the QS level when they cross the disk. To investigate the cause of this behavior, we perform a differential emission measure (DEM) analysis of the AIA snapshots using the DEM code developed by \citet{2015ApJ...807..143C}. Figure \ref{fig:dem} compares the EMs of different temperature bins for the two periods, when the sunspot (AR 12699: Section \ref{subsec:sunspot}) is behind the limb (2018 February 1: top) and when it is at the central meridian (2018 February 11: middle). When the spot is at the central meridian (in fact at the disk center), EMs are all enhanced in the spot region from the reference level (i.e., the QS period). Outside of the spot, however, the EM is reduced, especially for the temperature bins of $\log{T}=[5.75, 6.05]$.

To illustrate the EM reduction, the bottom left panel shows the azimuthally-averaged EM for $\log{T}=[5.75, 6.05]$ along the radial distance ($r$) from the disk center. As expected, the EM around the disk center ($r<130\arcsec$) is increased from the reference QS period. However, in contrast, the EM is reduced in the outskirts of the sunspot region ($130\arcsec <r<650\arcsec$). This ``dimming'' of the low temperature EMs survives even after the EMs are integrated over the full FOV of AIA (bottom right). This is the reason why the 171~{\AA} channel, which is most sensitive to the transition-region temperature at $\log{T}\sim 5.8$ (Table \ref{tab:aia}), is dimmed during the spot transit.

On the other hand, the EM is increased in all other temperature bins (i.e., $\log{T}\geq 6.1$) and, therefore, we can conjecture that the cause of the 171 dimming is the DEM shift to higher temperatures. This is also the reason for increased emissions in other AIA channels that probe the coronal plasmas (and the soft X-ray light curves).

\section{Summary and Discussion}\label{sec:summary}

\begin{figure}
\begin{center}
\includegraphics[width=0.45\textwidth]{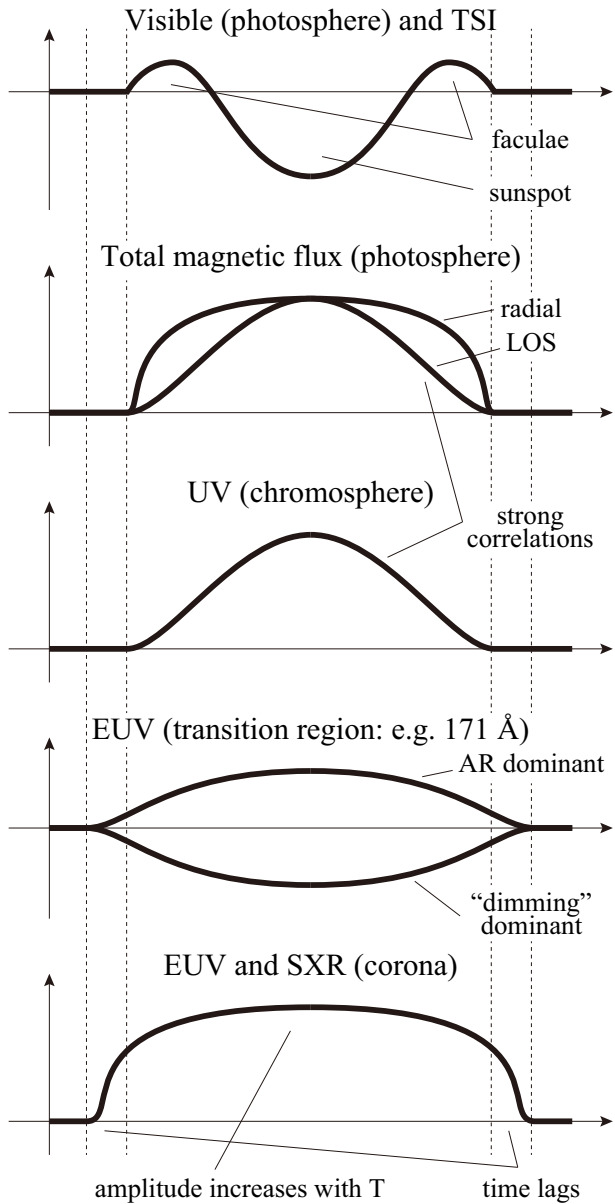}
\end{center}
\caption{Schematic illustration of transit AR light curves for different wavelengths. In each panel, the thick solid line shows the temporal variation of irradiance, while the reference QS level is indicated by the horizontal axis. The middle of the curves represents the time when the target AR is at the central meridian. This is the case for the transiting sunspot. The spotless plage lifts the ``dip'' in the visible and TSI curves to or even above the reference level, whereas the emerging flux makes all the curves asymmetric about the central meridian.\label{fig:summary}}
\end{figure}

In this study, we performed the analysis on the total and spectral solar irradiances, in particular for visible, UV, EUV, and soft X-rays, for the transiting events of an isolated sunspot, spotless plage, and emerging flux during the extremely quiet solar conditions that were selected from the past 10 years. The aim is to characterize the evolution of stellar light curves as starspots and other magnetic features rotate across the disks of other stars. The key findings of this study are summarized as follows and are schematically illustrated in Figure \ref{fig:summary}.

\begin{enumerate}
\item Visible continuum and TSI light curves become darker than the background level when the spot is around the central meridian because the darkening effect of umbra and penumbra dominates the brightening effect of faculae. On the contrary, the light curves are brighter when the spot is near the limb, as the faculae are contributing more. This behavior is consistent with previous TSI observations.

\item UV irradiances (e.g., 1600~{\AA} and 1700~{\AA}) take a mountain-shaped profile and correlate well with the LOS total unsigned magnetic flux in the photosphere. This is because these bandpasses are sensitive to the continuum emission and respond more to the presence of faculae than to the presence of spots.

\item EUV and soft X-ray irradiances show flat-top variations because the coronal plasma is optically thin and thus is less dependent on the viewing angle. This leads to the better correlations with the total radial magnetic flux than the LOS flux. In general, the amplitude of light curves are greater for higher characteristic temperatures. The soft X-ray flux in 1--8~{\AA} increases by more than an order of magnitude even outside of the solar flare periods.

\item EUV bands of the transition region ($\log{T}\lesssim 6.0$, e.g., 171~{\AA}) show in-phased or anti-phased evolutions with the other wavelengths. It becomes bright when the emission of the AR is dominant and darkened when the ``dimming'' effect of the surrounding regions is more contributing.

\item Magnetic features can influence irradiances for coronal lines even before they are evident in photospheric magnetograms. As a result, the presence of coronal loop tops (which are bright in EUV and soft X-ray wavelengths) can be seen from beyond the limb. The resulting enhancements in coronal light curves thus occur a few days before the corresponding changes in photospheric light curves as active regions rotate onto the solar disk. Likewise, EUV and soft X-ray light curves remain bright even after active regions have rotated off the disk, due to the vertical extent of the active region coronae. The length of time such AR coronae are visible above the limb likely depends on the heights of coronal loops above the photosphere, which are expected to be greater when the AR has a larger photospheric footprint.

\item Spotless plage regions produce light curves that are similar to the spot case in almost all aspects. The most remarkable difference is that the visible and TSI are not dimmed when the plage is near the central meridian, and instead remains at about the same level or even brighter than the background. The emerging flux case is similar to the spot case but all profiles become asymmetric about the central meridian.
\end{enumerate}

The goal of this study is to provide insight into the characteristics of the spectral irradiance on other solar-like stars. Apart from the near UV radiations (such as 1600~{\AA} and 1700~{\AA}), which might be used as the proxy for the total magnetic flux on the stellar surface like the \ion{Ca}{2} H and K monitoring, the time lags between the coronal and photospheric curves, for instance, are worth considering because they likely reflect the extension of the coronal magnetic fields. Investigating additional samples is therefore needed to better characterize the relationships between the time lags and the parameters of the transiting ARs.

Another feature of interest is the behavior of the transition region (e.g., 171~{\AA}) irradiance. The DEM analysis in Section \ref{subsec:171} revealed that when the sunspot crosses the disk, the ``dimming'' region is stretched far outside the spot to $650\arcsec$ and it has a cause in the DEM shift toward higher temperatures. This structure is reminiscent of the ``emerging dimming'', the 171~{\AA} intensity-reduced areas that are observed in the close proximity to the newly emerging flux regions for 3 to 14 hr \citep{2012ApJ...760L..29Z}. These authors ascribe the cause to the interaction between the newly emerging flux and the preexisting fields, i.e., plasma heating due to magnetic reconnection. Although the relation between the wider and longer-lived dimming considered here and the emerging dimmings is not clear, it is possible that the magnetic fields of an isolated spot group can extend far beyond and overwhelm the entire hemisphere \citep{2001ApJ...554..416P}. Perhaps similar physical mechanisms such as the heating and upflows caused by reconnection \citep{2010SoPh..263..105H} may be at play. The un-phased rotational modulation of these lines has not been well known on the Sun before, but at least \citet{2020arXiv200802702K} recently used {\it SDO}'s EUV Variability Experiment (EVE) and found that emissions of 171~{\AA} and other lines of $\log{T} \lesssim 6.0$ are similarly reduced when an AR transits across the solar disk. From the 171~{\AA} curves in the present study, it seems that younger ARs (sunspot and emerging flux) tend to produce the dimming behavior. If this is the case, we may be able to diagnose the plasma environments of the target starspots by comparing the coronal and transition-region light curves.

Here we only analyzed the spectral irradiance of isolated transiting objects, which are ideal and thus exceptionally rare even for our Sun in its activity minimum. Therefore, the conclusions found here should be applied only to the solar analogs, i.e., the slowly-rotating G-type stars. Nevertheless, the present study provides insight into how the AR atmospheres of other types of stars evolve in UV and X-rays. For instance, young ($\sim$100--500 Myr) rapidly-rotating solar analogs show flares with energies far exceeding the solar range (see references in Section \ref{sec:intro}). These stars show EM distributions with a characteristic peak around 10 MK, suggesting the possibility that their hotter coronae are maintained by frequent stellar flares \citep{2007JGRE..112.2008C,2007LRSP....4....3G}. The interaction of flare EUV radiations with the upper atmospheres of terrestrial-type exoplanets may drive atmospheric erosion and chemical reactions and thus is critical in planetary habitability \citep{2008SSRv..139..399L,2018A&ARv..26....2L,2014ApJ...790...57C,2016NatGe...9..452A,2017ApJ...836L...3A,2017NatSR...714141A,2020IJAsB..19..136A,2019A&A...624L..10J}. Furthermore, understanding the contributions of stellar ARs to spectral irradiance is critically important as such variations in light curves and inhomogeneities of stellar atmospheres can hamper the identification of terrestrial exoplanets around G- to M-type dwarfs \citep{2018arXiv180308708A,2020ApJ...890..121S}.

The measurements shown in this paper are entirely from near-Earth spacecraft, which are all located within the plane of the solar system. In stellar observations, however, the possibility of a star having multiple starspots and whose rotation axis is inclined to the LOS would need to be taken into account. To overcome these issues, we need to call for assistance of light curve modeling \citep[e.g.,][]{1988A&A...189..163S}. Future space UV missions have a potential to perform spectroscopic probing of stellar and exoplanetary atmospheres. To make most of these opportunities, further solar irradiance and spectroscopic studies are therefore necessary.

\acknowledgments

% referee
The authors wish to thank the anonymous referee for numerous helpful and constructive comments.
% data
Data are courtesy of the science teams of {\it SORCE}, {\it SDO}, and {\it Hinode}.
% SDO
HMI and AIA are instruments on board {\it SDO}, a mission for NASA's Living With a Star program.
% Hinode
{\it Hinode} is a Japanese mission developed and launched by ISAS/JAXA, with NAOJ as domestic partner and NASA and STFC (UK) as international partners. It is operated by these agencies in cooperation with ESA and NSC (Norway).
% Yang Liu, Xudong Sun, Yuta Notsu
The authors wish to thank Yang Liu for providing monthly HMI magnetogram animations, Xudong Sun for commenting on the noise estimation of HMI continuum data, and Yuta Notsu and Robert Stern for discussions on the stellar context.
% Toriumi
This work was supported by JSPS KAKENHI Grant Number 18H05234 and the NINS program for cross-disciplinary study (Grant Numbers 01321802 and 01311904) on Turbulence, Transport, and Heating Dynamics in Laboratory and Astrophysical Plasmas: ``SoLaBo-X''.
% Airapetian
The work of V.S.A. was supported by the NASA Exobiology grant 80NSSC17K0463, {\it TESS} Cycle 1 and {\it NICER} Cycle 2 GO programs.

%% To help institutions obtain information on the effectiveness of their 
%% telescopes the AAS Journals has created a group of keywords for telescope 
%% facilities.
%
%% Following the acknowledgments section, use the following syntax and the
%% \facility{} or \facilities{} macros to list the keywords of facilities used 
%% in the research for the paper.  Each keyword is check against the master 
%% list during copy editing.  Individual instruments can be provided in 
%% parentheses, after the keyword, but they are not verified.

\vspace{5mm}
%\facilities{}

%% Similar to \facility{}, there is the optional \software command to allow 
%% authors a place to specify which programs were used during the creation of 
%% the manuscript. Authors should list each code and include either a
%% citation or url to the code inside ()s when available.

\software{SolarSoftWare \citep{1998SoPh..182..497F}}

\bibliography{toriumi2020}{}

\begin{thebibliography}{}
\expandafter\ifx\csname natexlab\endcsname\relax\def\natexlab#1{#1}\fi

\bibitem[{{Abramenko} {et~al.}(2017){Abramenko}, {Kutsenko}, {Tikhonova}, \&
  {Yurchyshyn}}]{2017SoPh..292...48A}
{Abramenko}, V.~I., {Kutsenko}, A.~S., {Tikhonova}, O.~I., \& {Yurchyshyn},
  V.~B. 2017, \solphys, 292, 48

\bibitem[{{Airapetian} {et~al.}(2016){Airapetian}, {Glocer}, {Gronoff},
  {H{\'e}brard}, \& {Danchi}}]{2016NatGe...9..452A}
{Airapetian}, V.~S., {Glocer}, A., {Gronoff}, G., {H{\'e}brard}, E., \&
  {Danchi}, W. 2016, Nature Geoscience, 9, 452

\bibitem[{{Airapetian} {et~al.}(2017{\natexlab{a}}){Airapetian}, {Glocer},
  {Khazanov}, {Loyd}, {France}, {Sojka}, {Danchi}, \&
  {Liemohn}}]{2017ApJ...836L...3A}
{Airapetian}, V.~S., {Glocer}, A., {Khazanov}, G.~V., {et~al.}
  2017{\natexlab{a}}, \apjl, 836, L3

\bibitem[{{Airapetian} {et~al.}(2017{\natexlab{b}}){Airapetian}, {Jackman},
  {Mlynczak}, {Danchi}, \& {Hunt}}]{2017NatSR...714141A}
{Airapetian}, V.~S., {Jackman}, C.~H., {Mlynczak}, M., {Danchi}, W., \& {Hunt},
  L. 2017{\natexlab{b}}, Scientific Reports, 7, 14141

\bibitem[{{Airapetian} {et~al.}(2020){Airapetian}, {Barnes}, {Cohen},
  {Collinson}, {Danchi}, {Dong}, {Del Genio}, {France}, {Garcia-Sage},
  {Glocer}, {Gopalswamy}, {Grenfell}, {Gronoff}, {G{\"u}del}, {Herbst},
  {Henning}, {Jackman}, {Jin}, {Johnstone}, {Kaltenegger}, {Kay}, {Kobayashi},
  {Kuang}, {Li}, {Lynch}, {L{\"u}ftinger}, {Luhmann}, {Maehara}, {Mlynczak},
  {Notsu}, {Osten}, {Ramirez}, {Rugheimer}, {Scheucher}, {Schlieder},
  {Shibata}, {Sousa-Silva}, {Stamenkovi{\'c}}, {Strangeway}, {Usmanov},
  {Vergados}, {Verkhoglyadova}, {Vidotto}, {Voytek}, {Way}, {Zank}, \&
  {Yamashiki}}]{2020IJAsB..19..136A}
{Airapetian}, V.~S., {Barnes}, R., {Cohen}, O., {et~al.} 2020, International
  Journal of Astrobiology, 19, 136

\bibitem[{{Apai} {et~al.}(2018){Apai}, {Rackham}, {Giampapa}, {Angerhausen},
  {Teske}, {Barstow}, {Carone}, {Cegla}, {Domagal-Goldman}, {Espinoza},
  {Giles}, {Gully-Santiago}, {Haywood}, {Hu}, {Jordan}, {Kreidberg}, {Line},
  {Llama}, {L{\'o}pez-Morales}, {Marley}, \& {de Wit}}]{2018arXiv180308708A}
{Apai}, D., {Rackham}, B.~V., {Giampapa}, M.~S., {et~al.} 2018, arXiv e-prints,
  arXiv:1803.08708

\bibitem[{{Benz}(2017)}]{2017LRSP...14....2B}
{Benz}, A.~O. 2017, Living Reviews in Solar Physics, 14, 2

\bibitem[{{Berdyugina}(2005)}]{2005LRSP....2....8B}
{Berdyugina}, S.~V. 2005, Living Reviews in Solar Physics, 2, 8

\bibitem[{{Boerner} {et~al.}(2012){Boerner}, {Edwards}, {Lemen}, {Rausch},
  {Schrijver}, {Shine}, {Shing}, {Stern}, {Tarbell}, {Title}, {Wolfson},
  {Soufli}, {Spiller}, {Gullikson}, {McKenzie}, {Windt}, {Golub}, {Podgorski},
  {Testa}, \& {Weber}}]{2012SoPh..275...41B}
{Boerner}, P., {Edwards}, C., {Lemen}, J., {et~al.} 2012, \solphys, 275, 41

\bibitem[{{Carlsson} {et~al.}(2004){Carlsson}, {Stein}, {Nordlund}, \&
  {Scharmer}}]{2004ApJ...610L.137C}
{Carlsson}, M., {Stein}, R.~F., {Nordlund}, {\r{A}}., \& {Scharmer}, G.~B.
  2004, \apjl, 610, L137

\bibitem[{{Cheung} {et~al.}(2015){Cheung}, {Boerner}, {Schrijver}, {Testa},
  {Chen}, {Peter}, \& {Malanushenko}}]{2015ApJ...807..143C}
{Cheung}, M. C.~M., {Boerner}, P., {Schrijver}, C.~J., {et~al.} 2015, \apj,
  807, 143

\bibitem[{{Cnossen} {et~al.}(2007){Cnossen}, {Sanz-Forcada}, {Favata},
  {Witasse}, {Zegers}, \& {Arnold}}]{2007JGRE..112.2008C}
{Cnossen}, I., {Sanz-Forcada}, J., {Favata}, F., {et~al.} 2007, Journal of
  Geophysical Research (Planets), 112, E02008

\bibitem[{{Cohen} {et~al.}(2014){Cohen}, {Drake}, {Glocer}, {Garraffo},
  {Poppenhaeger}, {Bell}, {Ridley}, \& {Gombosi}}]{2014ApJ...790...57C}
{Cohen}, O., {Drake}, J.~J., {Glocer}, A., {et~al.} 2014, \apj, 790, 57

\bibitem[{{Couvidat} {et~al.}(2016){Couvidat}, {Schou}, {Hoeksema}, {Bogart},
  {Bush}, {Duvall}, {Liu}, {Norton}, \& {Scherrer}}]{2016SoPh..291.1887C}
{Couvidat}, S., {Schou}, J., {Hoeksema}, J.~T., {et~al.} 2016, \solphys, 291,
  1887

\bibitem[{{Flaccomio} {et~al.}(2005){Flaccomio}, {Micela}, {Sciortino},
  {Feigelson}, {Herbst}, {Favata}, {Harnden}, \&
  {Vrtilek}}]{2005ApJS..160..450F}
{Flaccomio}, E., {Micela}, G., {Sciortino}, S., {et~al.} 2005, \apjs, 160, 450

\bibitem[{{Fletcher} {et~al.}(2011){Fletcher}, {Dennis}, {Hudson}, {Krucker},
  {Phillips}, {Veronig}, {Battaglia}, {Bone}, {Caspi}, {Chen}, {Gallagher},
  {Grigis}, {Ji}, {Liu}, {Milligan}, \& {Temmer}}]{2011SSRv..159...19F}
{Fletcher}, L., {Dennis}, B.~R., {Hudson}, H.~S., {et~al.} 2011, \ssr, 159, 19

\bibitem[{{France} {et~al.}(2019){France}, {Fleming}, {Drake}, {Mason},
  {Youngblood}, {Bourrier}, {Fossati}, {Froning}, {Koskinen}, {Kruczek},
  {Lipscy}, {McEntaffer}, {Romaine}, {Siegmund}, \&
  {Wilkinson}}]{2019SPIE11118E..08F}
{France}, K., {Fleming}, B.~T., {Drake}, J.~J., {et~al.} 2019, in Society of
  Photo-Optical Instrumentation Engineers (SPIE) Conference Series, Vol. 11118,
  \procspie, 1111808

\bibitem[{{Freeland} \& {Handy}(1998)}]{1998SoPh..182..497F}
{Freeland}, S.~L., \& {Handy}, B.~N. 1998, \solphys, 182, 497

\bibitem[{{Golding} {et~al.}(2014){Golding}, {Carlsson}, \&
  {Leenaarts}}]{2014ApJ...784...30G}
{Golding}, T.~P., {Carlsson}, M., \& {Leenaarts}, J. 2014, \apj, 784, 30

\bibitem[{{Golub} {et~al.}(2007){Golub}, {Deluca}, {Austin}, {Bookbinder},
  {Caldwell}, {Cheimets}, {Cirtain}, {Cosmo}, {Reid}, {Sette}, {Weber},
  {Sakao}, {Kano}, {Shibasaki}, {Hara}, {Tsuneta}, {Kumagai}, {Tamura},
  {Shimojo}, {McCracken}, {Carpenter}, {Haight}, {Siler}, {Wright}, {Tucker},
  {Rutledge}, {Barbera}, {Peres}, \& {Varisco}}]{2007SoPh..243...63G}
{Golub}, L., {Deluca}, E., {Austin}, G., {et~al.} 2007, \solphys, 243, 63

\bibitem[{{G{\"u}del}(2007)}]{2007LRSP....4....3G}
{G{\"u}del}, M. 2007, Living Reviews in Solar Physics, 4, 3

\bibitem[{{Hagenaar}(2001)}]{2001ApJ...555..448H}
{Hagenaar}, H.~J. 2001, \apj, 555, 448

\bibitem[{{Harra} {et~al.}(2010){Harra}, {Magara}, {Hara}, {Tsuneta},
  {Okamoto}, \& {Wallace}}]{2010SoPh..263..105H}
{Harra}, L.~K., {Magara}, T., {Hara}, H., {et~al.} 2010, \solphys, 263, 105

\bibitem[{{Hudson} {et~al.}(1982){Hudson}, {Silva}, {Woodard}, \&
  {Willson}}]{1982SoPh...76..211H}
{Hudson}, H.~S., {Silva}, S., {Woodard}, M., \& {Willson}, R.~C. 1982,
  \solphys, 76, 211

\bibitem[{{Johnstone} {et~al.}(2019){Johnstone}, {Khodachenko},
  {L{\"u}ftinger}, {Kislyakova}, {Lammer}, \&
  {G{\"u}del}}]{2019A&A...624L..10J}
{Johnstone}, C.~P., {Khodachenko}, M.~L., {L{\"u}ftinger}, T., {et~al.} 2019,
  \aap, 624, L10

\bibitem[{{Kazachenko} \& {Hudson}(2020)}]{2020arXiv200802702K}
{Kazachenko}, M.~D., \& {Hudson}, H. 2020, arXiv e-prints, arXiv:2008.02702

\bibitem[{{Keller} {et~al.}(2004){Keller}, {Sch{\"u}ssler}, {V{\"o}gler}, \&
  {Zakharov}}]{2004ApJ...607L..59K}
{Keller}, C.~U., {Sch{\"u}ssler}, M., {V{\"o}gler}, A., \& {Zakharov}, V. 2004,
  \apjl, 607, L59

\bibitem[{{Kopp} {et~al.}(2005){Kopp}, {Lawrence}, \&
  {Rottman}}]{2005SoPh..230..129K}
{Kopp}, G., {Lawrence}, G., \& {Rottman}, G. 2005, \solphys, 230, 129

\bibitem[{{Kosugi} {et~al.}(2007){Kosugi}, {Matsuzaki}, {Sakao}, {Shimizu},
  {Sone}, {Tachikawa}, {Hashimoto}, {Minesugi}, {Ohnishi}, {Yamada}, {Tsuneta},
  {Hara}, {Ichimoto}, {Suematsu}, {Shimojo}, {Watanabe}, {Shimada}, {Davis},
  {Hill}, {Owens}, {Title}, {Culhane}, {Harra}, {Doschek}, \&
  {Golub}}]{2007SoPh..243....3K}
{Kosugi}, T., {Matsuzaki}, K., {Sakao}, T., {et~al.} 2007, \solphys, 243, 3

\bibitem[{{Lammer} {et~al.}(2008){Lammer}, {Kasting}, {Chassefi{\`e}re},
  {Johnson}, {Kulikov}, \& {Tian}}]{2008SSRv..139..399L}
{Lammer}, H., {Kasting}, J.~F., {Chassefi{\`e}re}, E., {et~al.} 2008, \ssr,
  139, 399

\bibitem[{{Lammer} {et~al.}(2018){Lammer}, {Zerkle}, {Gebauer}, {Tosi},
  {Noack}, {Scherf}, {Pilat-Lohinger}, {G{\"u}del}, {Grenfell}, {Godolt}, \&
  {Nikolaou}}]{2018A&ARv..26....2L}
{Lammer}, H., {Zerkle}, A.~L., {Gebauer}, S., {et~al.} 2018, \aapr, 26, 2

\bibitem[{{Lemen} {et~al.}(2012){Lemen}, {Title}, {Akin}, {Boerner}, {Chou},
  {Drake}, {Duncan}, {Edwards}, {Friedlaender}, {Heyman}, {Hurlburt}, {Katz},
  {Kushner}, {Levay}, {Lindgren}, {Mathur}, {McFeaters}, {Mitchell}, {Rehse},
  {Schrijver}, {Springer}, {Stern}, {Tarbell}, {Wuelser}, {Wolfson}, {Yanari},
  {Bookbinder}, {Cheimets}, {Caldwell}, {Deluca}, {Gates}, {Golub}, {Park},
  {Podgorski}, {Bush}, {Scherrer}, {Gummin}, {Smith}, {Auker}, {Jerram},
  {Pool}, {Soufli}, {Windt}, {Beardsley}, {Clapp}, {Lang}, \&
  {Waltham}}]{2012SoPh..275...17L}
{Lemen}, J.~R., {Title}, A.~M., {Akin}, D.~J., {et~al.} 2012, \solphys, 275, 17

\bibitem[{{Maehara} {et~al.}(2012){Maehara}, {Shibayama}, {Notsu}, {Notsu},
  {Nagao}, {Kusaba}, {Honda}, {Nogami}, \& {Shibata}}]{2012Natur.485..478M}
{Maehara}, H., {Shibayama}, T., {Notsu}, S., {et~al.} 2012, \nat, 485, 478

\bibitem[{{Namekata} {et~al.}(2019){Namekata}, {Maehara}, {Notsu}, {Toriumi},
  {Hayakawa}, {Ikuta}, {Notsu}, {Honda}, {Nogami}, \&
  {Shibata}}]{2019ApJ...871..187N}
{Namekata}, K., {Maehara}, H., {Notsu}, Y., {et~al.} 2019, \apj, 871, 187

\bibitem[{{Namekata} {et~al.}(2020){Namekata}, {Davenport}, {Morris}, {Hawley},
  {Maehara}, {Notsu}, {Toriumi}, {Ikuta}, {Notsu}, {Honda}, {Nogami}, \&
  {Shibata}}]{2020ApJ...891..103N}
{Namekata}, K., {Davenport}, J. R.~A., {Morris}, B.~M., {et~al.} 2020, \apj,
  891, 103

\bibitem[{{Narukage} {et~al.}(2011){Narukage}, {Sakao}, {Kano}, {Hara},
  {Shimojo}, {Bando}, {Urayama}, {Deluca}, {Golub}, {Weber}, {Grigis},
  {Cirtain}, \& {Tsuneta}}]{2011SoPh..269..169N}
{Narukage}, N., {Sakao}, T., {Kano}, R., {et~al.} 2011, \solphys, 269, 169

\bibitem[{{Notsu} {et~al.}(2013){Notsu}, {Shibayama}, {Maehara}, {Notsu},
  {Nagao}, {Honda}, {Ishii}, {Nogami}, \& {Shibata}}]{2013ApJ...771..127N}
{Notsu}, Y., {Shibayama}, T., {Maehara}, H., {et~al.} 2013, \apj, 771, 127

\bibitem[{{Notsu} {et~al.}(2019){Notsu}, {Maehara}, {Honda}, {Hawley},
  {Davenport}, {Namekata}, {Notsu}, {Ikuta}, {Nogami}, \&
  {Shibata}}]{2019ApJ...876...58N}
{Notsu}, Y., {Maehara}, H., {Honda}, S., {et~al.} 2019, \apj, 876, 58

\bibitem[{{O'Dwyer} {et~al.}(2010){O'Dwyer}, {Del Zanna}, {Mason}, {Weber}, \&
  {Tripathi}}]{2010A&A...521A..21O}
{O'Dwyer}, B., {Del Zanna}, G., {Mason}, H.~E., {Weber}, M.~A., \& {Tripathi},
  D. 2010, \aap, 521, A21

\bibitem[{{Pesnell} {et~al.}(2012){Pesnell}, {Thompson}, \&
  {Chamberlin}}]{2012SoPh..275....3P}
{Pesnell}, W.~D., {Thompson}, B.~J., \& {Chamberlin}, P.~C. 2012, \solphys,
  275, 3

\bibitem[{{Pevtsov} \& {Acton}(2001)}]{2001ApJ...554..416P}
{Pevtsov}, A.~A., \& {Acton}, L.~W. 2001, \apj, 554, 416

\bibitem[{{Redfield} {et~al.}(2006){Redfield}, {Linsky}, {Ayres}, {Brown}, \&
  {Herczeg}}]{2006ASPC..348..269R}
{Redfield}, S., {Linsky}, J.~L., {Ayres}, T.~R., {Brown}, A., \& {Herczeg},
  G.~J. 2006, in Astronomical Society of the Pacific Conference Series, Vol.
  348, Astrophysics in the Far Ultraviolet: Five Years of Discovery with FUSE,
  ed. G.~{Sonneborn}, H.~W. {Moos}, \& B.~G. {Andersson}, 269

\bibitem[{{Rottman}(2005)}]{2005SoPh..230....7R}
{Rottman}, G. 2005, \solphys, 230, 7

\bibitem[{{Sanz-Forcada} {et~al.}(2003){Sanz-Forcada}, {Maggio}, \&
  {Micela}}]{2003A&A...408.1087S}
{Sanz-Forcada}, J., {Maggio}, A., \& {Micela}, G. 2003, \aap, 408, 1087

\bibitem[{{Scherrer} {et~al.}(2012){Scherrer}, {Schou}, {Bush}, {Kosovichev},
  {Bogart}, {Hoeksema}, {Liu}, {Duvall}, {Zhao}, {Title}, {Schrijver},
  {Tarbell}, \& {Tomczyk}}]{2012SoPh..275..207S}
{Scherrer}, P.~H., {Schou}, J., {Bush}, R.~I., {et~al.} 2012, \solphys, 275,
  207

\bibitem[{{Schou} {et~al.}(2012){Schou}, {Scherrer}, {Bush}, {Wachter},
  {Couvidat}, {Rabello-Soares}, {Bogart}, {Hoeksema}, {Liu}, {Duvall}, {Akin},
  {Allard}, {Miles}, {Rairden}, {Shine}, {Tarbell}, {Title}, {Wolfson},
  {Elmore}, {Norton}, \& {Tomczyk}}]{2012SoPh..275..229S}
{Schou}, J., {Scherrer}, P.~H., {Bush}, R.~I., {et~al.} 2012, \solphys, 275,
  229

\bibitem[{{Schrijver}(1988)}]{1988A&A...189..163S}
{Schrijver}, C.~J. 1988, \aap, 189, 163

\bibitem[{{Schrijver}(2009)}]{2009AdSpR..43..739S}
---. 2009, Advances in Space Research, 43, 739

\bibitem[{{Schrijver}(2020)}]{2020ApJ...890..121S}
---. 2020, \apj, 890, 121

\bibitem[{{Shibata} \& {Magara}(2011)}]{2011LRSP....8....6S}
{Shibata}, K., \& {Magara}, T. 2011, Living Reviews in Solar Physics, 8, 6

\bibitem[{{Shibayama} {et~al.}(2013){Shibayama}, {Maehara}, {Notsu}, {Notsu},
  {Nagao}, {Honda}, {Ishii}, {Nogami}, \& {Shibata}}]{2013ApJS..209....5S}
{Shibayama}, T., {Maehara}, H., {Notsu}, S., {et~al.} 2013, \apjs, 209, 5

\bibitem[{{Shustov} {et~al.}(2014){Shustov}, {G{\'o}mez de Castro}, {Sachkov},
  {Moisheev}, {Kanev}, {L{\'o}pez-Santiago}, {Malkov}, {Nasonov}, {Bel{\'e}n
  Perea}, {S{\'a}nchez}, {Savanov}, {Shugarov}, {Sichevskiy}, {Vlasenko}, \&
  {Ya{\~n}ez}}]{2014Ap&SS.354..155S}
{Shustov}, B., {G{\'o}mez de Castro}, A.~I., {Sachkov}, M., {et~al.} 2014,
  \apss, 354, 155

\bibitem[{{Sim{\~o}es} {et~al.}(2019){Sim{\~o}es}, {Reid}, {Milligan}, \&
  {Fletcher}}]{2019ApJ...870..114S}
{Sim{\~o}es}, P. J.~A., {Reid}, H. A.~S., {Milligan}, R.~O., \& {Fletcher}, L.
  2019, \apj, 870, 114

\bibitem[{{Solanki}(2003)}]{2003A&ARv..11..153S}
{Solanki}, S.~K. 2003, \aapr, 11, 153

\bibitem[{{Solanki} {et~al.}(2013){Solanki}, {Krivova}, \&
  {Haigh}}]{2013ARA&A..51..311S}
{Solanki}, S.~K., {Krivova}, N.~A., \& {Haigh}, J.~D. 2013, \araa, 51, 311

\bibitem[{{Spruit}(1976)}]{1976SoPh...50..269S}
{Spruit}, H.~C. 1976, \solphys, 50, 269

\bibitem[{{Stern} \& {Skumanich}(1983)}]{1983ApJ...267..232S}
{Stern}, R.~A., \& {Skumanich}, A. 1983, \apj, 267, 232

\bibitem[{{Stern} {et~al.}(1992){Stern}, {Uchida}, {Walter}, {Vilhu},
  {Hannikainen}, {Brown}, {Veale}, \& {Haisch}}]{1992ApJ...391..760S}
{Stern}, R.~A., {Uchida}, Y., {Walter}, F., {et~al.} 1992, \apj, 391, 760

\bibitem[{{Strassmeier}(2009)}]{2009A&ARv..17..251S}
{Strassmeier}, K.~G. 2009, \aapr, 17, 251

\bibitem[{{Takeda} {et~al.}(2016){Takeda}, {Yoshimura}, \&
  {Saar}}]{2016SoPh..291..317T}
{Takeda}, A., {Yoshimura}, K., \& {Saar}, S.~H. 2016, \solphys, 291, 317

\bibitem[{{Toriumi} {et~al.}(2014){Toriumi}, {Hayashi}, \&
  {Yokoyama}}]{2014ApJ...794...19T}
{Toriumi}, S., {Hayashi}, K., \& {Yokoyama}, T. 2014, \apj, 794, 19

\bibitem[{{Toriumi} {et~al.}(2013){Toriumi}, {Iida}, {Bamba}, {Kusano},
  {Imada}, \& {Inoue}}]{2013ApJ...773..128T}
{Toriumi}, S., {Iida}, Y., {Bamba}, Y., {et~al.} 2013, \apj, 773, 128

\bibitem[{{Toriumi} {et~al.}(2017){Toriumi}, {Schrijver}, {Harra}, {Hudson}, \&
  {Nagashima}}]{2017ApJ...834...56T}
{Toriumi}, S., {Schrijver}, C.~J., {Harra}, L.~K., {Hudson}, H., \&
  {Nagashima}, K. 2017, \apj, 834, 56

\bibitem[{{Toriumi} \& {Wang}(2019)}]{2019LRSP...16....3T}
{Toriumi}, S., \& {Wang}, H. 2019, Living Reviews in Solar Physics, 16, 3

\bibitem[{{Warren}(2005)}]{2005ApJS..157..147W}
{Warren}, H.~P. 2005, \apjs, 157, 147

\bibitem[{{Willson} {et~al.}(1981){Willson}, {Gulkis}, {Janssen}, {Hudson}, \&
  {Chapman}}]{1981Sci...211..700W}
{Willson}, R.~C., {Gulkis}, S., {Janssen}, M., {Hudson}, H.~S., \& {Chapman},
  G.~A. 1981, Science, 211, 700

\bibitem[{{Withbroe}(2006)}]{2006SoPh..235..369W}
{Withbroe}, G.~L. 2006, \solphys, 235, 369

\bibitem[{{Woods} {et~al.}(2004){Woods}, {Eparvier}, {Fontenla}, {Harder},
  {Kopp}, {McClintock}, {Rottman}, {Smiley}, \& {Snow}}]{2004GeoRL..3110802W}
{Woods}, T.~N., {Eparvier}, F.~G., {Fontenla}, J., {et~al.} 2004, \grl, 31,
  L10802

\bibitem[{{Zahid} {et~al.}(2004){Zahid}, {Hudson}, \&
  {Fr{\"O}hlich}}]{2004SoPh..222....1Z}
{Zahid}, H.~J., {Hudson}, H.~S., \& {Fr{\"O}hlich}, C. 2004, \solphys, 222, 1

\bibitem[{{Zhang} {et~al.}(2012){Zhang}, {Yang}, {Liu}, \&
  {Sun}}]{2012ApJ...760L..29Z}
{Zhang}, J., {Yang}, S., {Liu}, Y., \& {Sun}, X. 2012, \apjl, 760, L29

\end{thebibliography}
\bibliographystyle{aasjournal}

%% This command is needed to show the entire author+affiliation list when
%% the collaboration and author truncation commands are used.  It has to
%% go at the end of the manuscript.
%\allauthors

%% Include this line if you are using the \added, \replaced, \deleted
%% commands to see a summary list of all changes at the end of the article.
%\listofchanges

\end{document}